\documentclass[useAMS,usenatbib,usegraphicx]{mn2e}
\begin{document}
%
\title[Low $T/|W|$ dynamical instability] 
{
Low $T/|W|$ dynamical instability in differentially rotating stars:
Diagnosis with canonical angular momentum}
%
\author[M. Saijo and S. Yoshida]
{Motoyuki Saijo$^{1}$
\thanks{E-mail:~ms1@maths.soton.ac.uk~(MS);
  yoshida@physics.fau.edu~(SY)}
\thanks{
Present address: School of Mathematics,
University of Southampton,
Southampton SO17 1BJ.
}
and Shin'ichirou Yoshida$^{2}$
\footnotemark[1]
\thanks{
Present address: Department of Physics, Florida Atlantic University,
Boca Raton, FL 33431, USA.}\\
$^{1}$ Laboratoire de l'Univers et de ses Th\'eories,
Observatoire de Paris, F-92195 Meudon Cedex, France\\
$^{2}$ Department of Physics, University of Wisconsin-Milwaukee,
Milwaukee, WI 53211, USA}
%
\date{Accepted 2006 February 20. Received 2006 February 13; 
  in original form 2005 May 25}
%
\volume{in press}
\pagerange{\pageref{firstpage}--\pageref{lastpage}}
\pubyear{2006}
%
\eqsecnum
\maketitle
\label{firstpage}
\begin{abstract}
We study the nature of non-axisymmetric dynamical instabilities in  
differentially rotating stars with both linear eigenmode analysis and
hydrodynamic simulations in Newtonian gravity.  We especially
investigate the following three types of instability; the one-armed
spiral instability, the low $T/|W|$ bar instability, and the high
$T/|W|$ bar instability, where $T$ is the rotational kinetic energy
and $W$ is the gravitational potential energy.  The nature of the
dynamical instabilities is clarified by using a canonical angular
momentum as a diagnostic.  We find that the one-armed spiral and the
low $T/|W|$ bar instabilities occur around the corotation radius, and
they grow through the inflow of canonical angular momentum around the
corotation radius.  The result is a clear contrast to that of a
classical dynamical bar instability in high $T/|W|$.  We also discuss
the feature of gravitational waves generated from these three types of
instability. 
\end{abstract}

\section{Introduction}
Stars in nature are usually rotating and may be subject to
non-axisymmetric rotational instabilities.  An analytically exact
treatment of these instabilities in linearized theory exists only for
incompressible equilibrium fluids in Newtonian gravity
\citep[e.g.,][]{Chandra69,Tassoul78,ST83}.  For these configurations,
global rotational instabilities may arise from non-radial toroidal
modes $e^{im\varphi}$ (where $m=\pm 1,\pm 2, \dots$ and $\varphi$ is
the azimuthal angle). 

For sufficiently rapid rotation, the $m=2$ bar mode becomes either 
{\em secularly} or {\em dynamically} unstable.  The onset of
instability can typically be marked by a critical value of the
dimensionless parameter $\beta \equiv T/|W|$, where $T$ is the
rotational kinetic energy and $W$ the gravitational potential energy.
Uniformly rotating, incompressible stars in Newtonian theory are
secularly unstable to bar-mode formation when $\beta \ga \beta_{\rm
  sec} \simeq 0.14$.  This instability can grow only in the presence
of some dissipative mechanism, like viscosity or gravitational
radiation, and the associated growth time-scale is the dissipative
time-scale, which is usually much longer than the dynamical time-scale
of the system.  By contrast, a dynamical instability to bar formation
sets in when $\beta \ga \beta_{\rm dyn} \simeq 0.27$.  This
instability is present independent of any dissipative mechanism, and 
the growth time is the hydrodynamic time-scale.

In addition to the classical dynamical instability mentioned above,
there have been several studies indicating that a dynamical
instability of the rotating stars occurs at low $T/|W|$, which is far
below the classical criterion of dynamical instability in Newtonian
gravity.  \citet{TH90} find in the self-gravitating ring and tori that
an $m=2$ dynamical instability occurs around $T/|W| \sim 0.16$ in the
lowest case in the centrally condensed configurations.  For rotating
stellar models, \citet*{SKE02,SKE03} find that $m=2$ dynamical
instability occurs around $T/|W| \sim O(10^{-2})$ for a high degree
($\Omega_{\rm c} / \Omega_{\rm s} \approx 10$) of differential
rotation.  Note that $\Omega_{\rm c}$ and $\Omega_{\rm s}$ are the
angular velocity at the centre and at the equatorial surface,
respectively.  \citet{CNLB01} found dynamical $m=1$ instability around
$T/|W| \sim 0.09$ in the $N = 3.33$ polytropic `toroidal' star with a
high degree ($\Omega_{\rm c} / \Omega_{\rm s} = 26$) of differential
rotation, and \citet*{SBS03} extended the results of dynamical $m=1$
instability to the cases with polytropic index $N \ga 2.5$ and
$\Omega_{\rm c} / \Omega_{\rm s} \ga 10$.

Computation of the onset of the dynamical instability, as well as the  
subsequent evolution of an unstable star, performed in a fully
nonlinear hydrodynamic simulation in Newtonian gravity,
\citep*[e.g.][]{TDM, DGTB86, WT, HCS, SHC, HC, PDD, TIPD, NCT} have
shown that $\beta_{\rm dyn}$ depends only very weakly on the stiffness
of the equation of state.  Once a bar has developed, the formation of
a two-arm spiral plays an important role in redistributing the angular 
momentum and forming a core-halo structure.  $\beta_{\rm dyn}$ are
smaller for stars with high degree of differential rotation
\citep{TH90, PDD, SKE02, SKE03}.  Simulations in relativistic
gravitation \citep*{SBS00,SSBS} have shown that $\beta_{\rm dyn}$
decreases with the compaction of the star, indicating that
relativistic gravitation enhances the bar mode instability.  

Recently, several studies have been focused on the collapse of the
rotating stars leading to non-axisymmetric dynamical instabilities in
three-dimensional hydrodynamics.  \citet*{DSY04} investigated the
collapse of a differentially rotating $N=1$ polytropic star in full
general relativity by depleting the pressure and found that the
collapsing star forms a torus which fragments into nonaxisymmetric
clumps.  \citet{SS05} investigated rotational core collapse in full
general relativity and found that a burst type of gravitational waves 
was emitted.  In addition, they argued that a very limited window for
the rotating star satisfies to exceed the threshold of dynamical
instability in the core collapse.  \citet{Saijo05} studied the 
spheroidal and toroidal configuration of the collapsing star in
conformally flat gravity, and found that toroidal configuration has a
potential source of gravitational waves due to the enhancement of the
non-axisymmetric dynamical instability.  \citet{Zink05} presented a
fragmentation of an $N=3$ toroidal polytropic star to both one-armed
spiral and a binary system in full general relativity, depending on
the type of initial density perturbation.  There the authors confirm
that the instabilities found in Newtonian gravity also appear in
general relativistic stars of astrophysical relevance.  \citet{Ott05}
performed gravitational collapse of unstable iron cores at the centre
of evolved massive stars in Newtonian gravity.  Their simulations
contained the evolutions from implosion of iron core (the computations
done in two-dimensional code) up to the post bounce phase, in which
they found growth of unstable $m=1$, $2$ oscillations.

One of the remarkable features of these low $T/|W|$ instabilities is
an appearance of the corotation modes.  As it is pointed out by
\citet*{WAJ05} the low $T/|W|$ unstable oscillation of bar-typed one
found by \citet{SKE02, SKE03} has a corotation point.  Here,
corotation means the pattern speed of oscillation in the azimuthal
direction coincides with a local rotational angular velocity of the
star.  It is well-known in the context of stellar or gaseous disk
system that the corotation of oscillation may lead to instabilities.
For instance, there have been several density wave models proposed to
explain spiral pattern in galaxies, in which wave amplification at the
corotation radius of spiral pattern is a key issue \citep{Shu92}.
Another example of importance of corotation is found in the theory of
thick disk (torus) around black holes.  Initiated by a discovery of a
dynamical instability of geometrically thick disk by
\citet{PP84, PP85, PP87}, several authors have studied these 
instabilities \citep*{Blaes85a, Blaes85b, Drury85, Blaes86, GGN86,
  Kojima86, Kojima89, Glatzel87a, Glatzel87b, GN88}.  Instabilities of 
these systems are thought not to be unique in their origin and in
their characteristics.  Some seem to be related to local shear of flow 
and to share a nature with Kelvin-Helmholtz instability.  Others may
be related to corotation of oscillation modes with averaged flow on
which the oscillation is present.  The mechanisms of instabilities by
corotation, however, seem not unique.  As is reminiscent to `Landau
amplification' of plasma wave \citep{Stix92}, a resonant interaction
of corotating wave with the background flow (in the case of Landau
amplification, background flow is that of charged particles) may
amplify the wave, by direct pumping of energy from background flow to
the oscillation.  The other may be an overreflection of waves at the
corotation which may be seen in waves propagating towards shear layer
\citep{Acheson76}.

The main purpose of this paper, in contrast to the preceding studies
of this issue, is to investigate the nature of low $T/|W|$ dynamical
instabilities, especially to study the qualitative difference of them
from the classical bar instability.  As is mentioned above, recent
studies have shown that dynamical instabilities are possible for
different region of the parameter space of rotating stars.  Observing
the existence of dynamical instabilities whose critical $T/|W|$ value
are well below the classical criterion of bar instability, it is
natural to raise a question on whether these two types, `high $T/|W|$'
and `low $T/|W|$', of dynamical instability are categorized in the
same type of dynamical instability or not.

Our study is done with both eigenmode analysis and hydrodynamical
analysis.  A non-linear hydrodynamical simulation is indispensable for
investigation of evolutionary process and final outcome of
instability, such as bar formation and spiral structure formation.
The nature of instability as a source of gravitational wave, which
interests us most, is only accessible through non-linear
hydrodynamical computations.  On the other hand, a linear eigenmode
analysis is in general easier to approach the dynamical instability of
a given equilibrium and it may be helpful to have physical insight on
the mechanism and the origin of the instability.  Therefore, a linear
eigenmode analysis and a non-linear simulation are complementary to
each other and they both help us to understand the nature of dynamical
instability.

As a simplified system mimicking the physical nature of the
differentially rotating fluid, we choose to study self-gravitating
cylinder models.  They have been used to study general dynamical
nature of such gaseous masses as stars, accretion disks and of stellar
system as galaxies.  Although there is no infinite-length cylinder in
the real world, it is expected to share some qualitative similarities
with realistic astrophysical objects \citep*{Ostriker65, Robe79,
  Balbinski85, IA85, IA86, Blaes86, Glatzel87a, Glatzel87b, Luyten88,
  Luyten89, Luyten90, CSS03}.  Especially it has served as a useful
model to investigate secular and dynamical instabilities of rotating
masses.  These works took advantage of a simple configuration of a
cylinder compared with a spheroid.

This paper is organized as follows.  In Sections \ref{sec:Nhydro} and
\ref{sec:LinearAnalysis} we present our hydrodynamical and eigenmode
analysis results of dynamical one-armed spiral and dynamical bar
instability.  We present our diagnosis of dynamical $m=1$ and $m=2$
instabilities by using a canonical angular momentum in Section
\ref{sec:Canonical}, and summarize our findings in Section
\ref{sec:Discussion}.  Throughout this paper we use gravitational
units with $G = 1$.  Latin indices run over spatial coordinates.

\section{Hydrodynamic simulations in Differentially Rotating Stars}
\label{sec:Nhydro}

Here, we briefly describe the basic equation of the perfect fluid
hydrodynamics in Newtonian gravity.  We follow \citet{SBS03} to
perform our $3$ dimensional Newtonian hydrodynamics assuming an
adiabatic $\Gamma$-law equation of state 
\begin{equation}
P = ( \Gamma - 1 ) \rho \varepsilon,
\label{eqn:GammaLaw}
\end{equation}
where $P$ is the pressure, $\Gamma$ the adiabatic index, $\rho$ the
mass density and $\varepsilon$ the specific internal energy density.
For perfect fluids, the Newtonian equations of hydrodynamics then
consist of the continuity equation
\begin{equation}
\frac{\partial \rho}{\partial t}
+\frac{\partial (\rho v^{i})}{\partial x^{i}} = 0,
\label{eqn:continuity}
\end{equation}
the energy equation
\begin{equation}
\frac{\partial e}{\partial t}+
\frac{\partial (e v^{j})}{\partial x^{j}} =
- \frac{1}{\Gamma} e^{-(\Gamma-1)} P_{\rm vis}
\frac{\partial v^{i}}{\partial x^{i}}
,
\end{equation}
and the Euler equation
\begin{equation}
\frac{\partial(\rho v_{i})}{\partial t}
+ \frac{\partial (\rho v_{i} v^{j})}{\partial x^{j}}
=
- \frac{\partial (P + P_{\rm vis})}{\partial x^{i}}
- \rho \frac{\partial \Phi}{\partial x^{i}}.
\end{equation}
Here $v^i$ is the fluid velocity, $\Phi$ is the gravitational
potential 
\begin{equation}
\triangle \Phi = 4 \pi \rho,
\end{equation}
and $e$ is defined according to
\begin{equation}
e = (\rho \varepsilon)^{1/\Gamma}.
\end{equation}
We use the same type of artificial viscosity pressure $P_{\rm vis}$ in  
\citet{SBS03}.  When evolving the above equations we limit the
stepsize $\Delta t$ by an appropriately chosen dynamical time.

We construct differentially rotating equilibrium stars based on 
\citet{Hachisu86}.  We assume a polytropic equation of state only to
construct an equilibrium star as
\begin{equation}
P = \kappa \rho^{1+1/N},
\end{equation}
where $\kappa$ is a constant, $N$ is the polytropic index.  We also
assume the `$j$-constant' rotation law, which has an exact meaning in
the limit of $d \rightarrow 0$, of the rotating stars
\begin{equation}
\Omega = \frac{j_{0}}{d^{2} + \varpi^{2}},
\label{eqn:omega}
\end{equation}
where $\Omega$ is the angular velocity, $j_{0}$ is a constant
parameter with units of specific angular momentum, and $\varpi$ is the
cylindrical radius.  The parameter $d$ determines the length scale
over which $\Omega$ changes; uniform rotation is achieved in the limit
$d \rightarrow \infty$, with keeping the ratio $j_0/d^2$ finite.  We
choose the same data sets as \citet{SBS03} for investigating low
$T/|W|$ dynamical instabilities in differentially rotating stars
(models I and II in Table~\ref{tab:initial} corresponds to Tables II
and I of \citet{SBS03}, respectively).  We also construct an
equilibrium star with weak differential rotation in high $T/|W|$,
which excites the standard dynamical bar instability,
\citep[e.g.,][]{Chandra69}.  The characteristic parameters are
summarized in Table~\ref{tab:initial}.

To enhance any $m=1$ or $m=2$ instability, we disturb the initial
equilibrium mass density $\rho_{\rm eq}$ by a non-axisymmetric
perturbation according to 
\begin{equation}
\rho = \rho_{\rm eq}
\left( 1 +
  \delta^{(1)} \frac{x+y}{R_{\rm eq}} +
  \delta^{(2)} \frac{x^{2}-y^{2}}{R_{\rm eq}^{2}}
\right),
\label{eqn:DPerturb}
\end{equation}
where $R_{\rm eq}$ is the equatorial radius, with $\delta^{(1)} =
\delta^{(2)} \approx 1.7 - 2.8 \times 10^{-3}$ in all our simulations.
We also introduce `dipole' $D$ and `quadrupole' $Q$ diagnostics to
monitor the development of $m=1$ and $m=2$ modes as 
\begin{eqnarray}
D &=& \left< e^{i m \varphi} \right>_{m=1} =
\frac{1}{M} \int \rho \frac{x + i y}{\varpi} dV,
\\
Q &=& \left< e^{i m \varphi} \right>_{m=2} =
\frac{1}{M} \int \rho \frac{(x^{2}-y^{2}) + i (2 x y)}{\varpi^{2}} dV,
\nonumber \\
\end{eqnarray}
respectively.

We compute approximate gravitational waveforms by using the quadrupole  
formula.  In the radiation zone, gravitational waves can be described
by a transverse-traceless, perturbed metric $h_{ij}^{TT}$ with respect
to flat spacetime.  In the quadrupole formula, $h_{ij}^{TT}$ is found
from \citep*{MTW} 
\begin{equation}
h_{ij}^{TT}= \frac{2}{r} \frac{d^{2}}{d t^{2}} I_{ij}^{TT},
\label{eqn:wave1}
\end{equation}
where $r$ is the distance to the source, $I_{ij}$ the quadrupole
moment of the mass distribution, and where $TT$ denotes the
transverse-traceless projection.  Choosing the direction of the wave
propagation to be along the rotational axis ($z$-axis), the two
polarization modes of gravitational waves can be determined from
\begin{equation}
h_{+} \equiv \frac{1}{2} (h_{xx}^{TT} - h_{yy}^{TT}),
\quad 
h_{\times} \equiv h_{xy}^{TT}.
\end{equation}
For observers along the rotation axis, we thus have
\begin{eqnarray}
\frac{r h_{+}}{M} &=&
\frac{1}{2 M} \frac{d^{2}}{d t^{2}} ( I_{xx}^{TT} - I_{yy}^{TT}),
\label{h+} 
\\
\frac{r h_{\times}}{M} &=&
\frac{1}{M} \frac{d^{2}}{d t^{2}} I_{xy}^{TT} 
\label{h-}
.
\end{eqnarray}

\begin{table*}
\centering
\begin{minipage}{15cm}
\caption
{Three differentially rotating equilibrium stars that trigger
  dynamical instability.
\label{tab:initial}}
\begin{tabular}{@{}ccccccccc@{}}
\hline
Model &
$N^{a}$ &
${d / R_{\rm eq}}^{b}$ &
${R_{\rm p} / R_{\rm eq}}^{c}$ &
${\Omega_{\rm c} / \Omega_{\rm s}}^{d}$ &
${\rho_{\rm c} / \rho_{\rm max}}^{e}$ &
${R_{\rm maxd}/R_{\rm eq}}^{f}$ &
${T/|W|}^{g}$ &
Dominant unstable mode
\\
\hline
I & $3.33$ &
$0.20$ & $0.413$ & $26.0$ & $0.491$ & $0.198$ &
$0.146$ & $m=1$
\\
II & $1.00$ &
$0.20$ & $0.250$ & $26.0$ & $0.160$ & $0.383$ &
$0.119$ & $m=2$
\\
III & $1.00$ &
$1.00$ & $0.250$ & $2.0$ & $0.837$ & $0.388$ &
$0.277$ & $m=2$
\\
\hline
\end{tabular}

\medskip
${}^{a} N$: polytropic index.
${}^{b} R_{\rm eq}$: equatorial radius.
${}^{c} R_{\rm pl}$: polar radius.
${}^{d} \Omega_{\rm c}$: central angular velocity;
$\Omega_{\rm s}$: equatorial surface angular velocity.
${}^{e} \rho_{\rm c}$: central mass density; 
$\rho_{\rm max}$: maximum mass density.
${}^{f} R_{\rm maxd}$: radius of maximum density.
${}^{g} T$: rotational kinetic energy;
$W$: gravitational binding energy.
\end{minipage}
\end{table*}

\begin{figure}
\centering
\includegraphics[keepaspectratio=true,width=8cm]{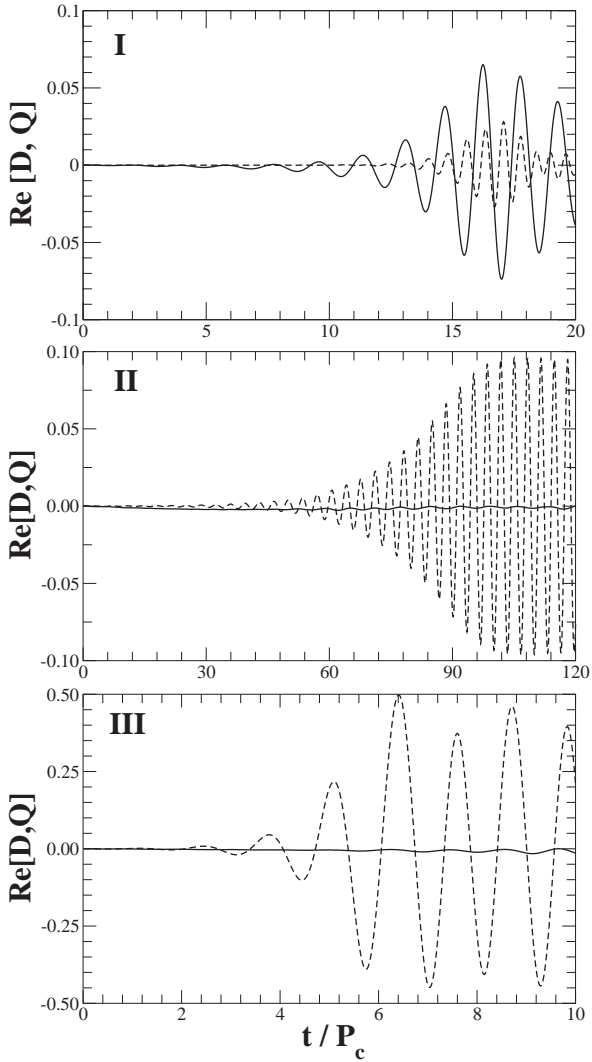}
\caption{
Diagnostics $D$ and $Q$ as a function of $t/P_{\rm c}$ for three
differentially rotating stars (see Table~\ref{tab:initial}).  Solid
and dotted lines denote the values of $D$ and $Q$, respectively.  The
Roman numeral in each panel corresponds to the model of the
differentially rotating stars, respectively.  Hereafter $P_{\rm c}$
represents the central rotation period.
}
\label{fig:dig}
\end{figure}

The time evolutions of the dipole diagnostic and the quadrupole
diagnostic are plotted in Fig.~\ref{fig:dig}.  We determine that the
system is stable to $m=1$ ($m=2$) mode when the dipole (quadrupole)
diagnostic remains small throughout the evolution, while the system is
unstable when the diagnostic grows exponentially at the early stage of
the evolution.  It is clearly seen in Fig.~\ref{fig:dig} that the star
is more unstable to the one-armed spiral mode for model I, and more
unstable to the bar mode for models II and III.  In fact, both
diagnostics grow for model I.  The dipole diagnostic, however, grows
larger than the quadrupole diagnostic, indicating that the $m=1$ mode
is the dominant unstable mode.

\begin{figure}
\centering
\includegraphics[keepaspectratio=true,width=8cm]{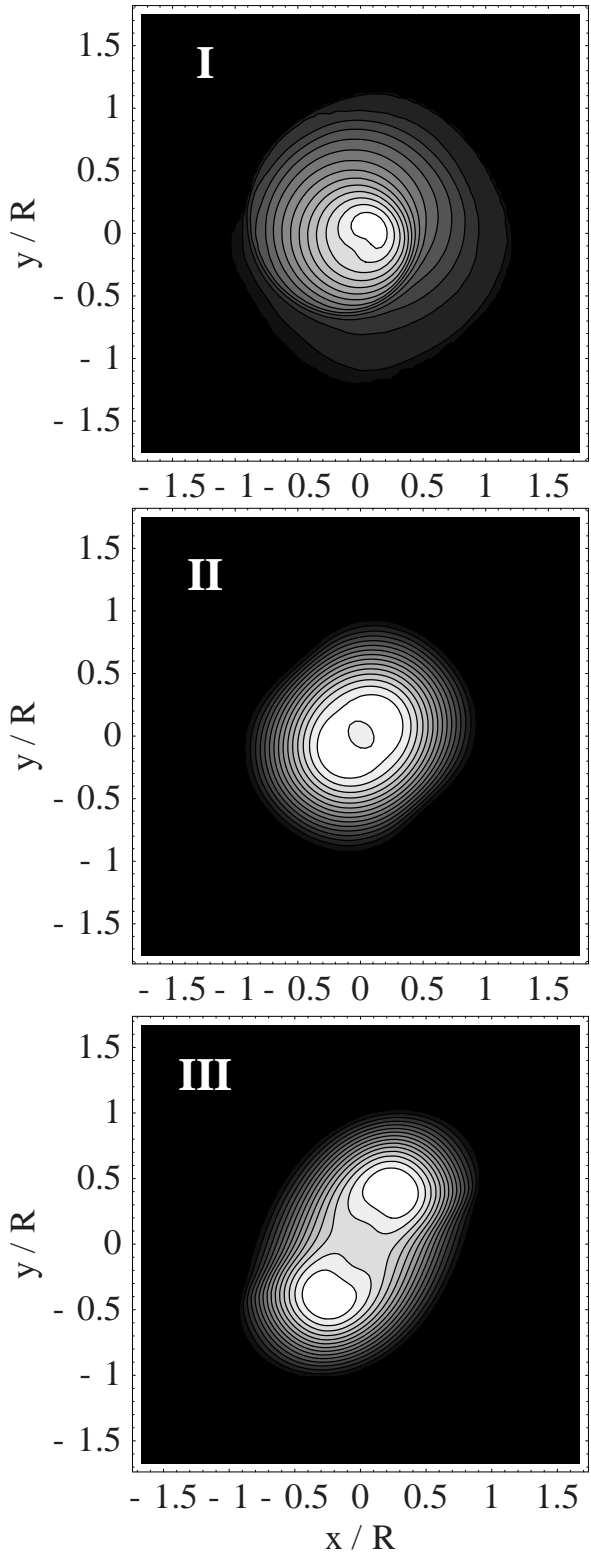}
\caption{
Density contours in the equatorial plane for three differentially rotating
stars (see Table~\ref{tab:initial}).  Models~I, II, and III are
plotted at the parameter ($t/P_{\rm c}$, $\rho_{\rm max} / \rho_{\rm
  max}^{(0)}$) = ($16.2$, $3.63$), ($134$, $1.26$), and ($5.49$,
$1.20$), where $\rho_{\rm max}$ is the maximum density, $\rho_{\rm
  max}^{(0)}$ is the maximum density at $t=0$, and $R$ is the
equatorial radius at $t=0$.  The contour lines denote densities $\rho
/ \rho_{\rm max} = 10^{- (16-i) \times 0.287} (i=1, \cdots, 15)$ for
model~I and $\rho / \rho_{\rm max} = 6.67 (16-i) \times 10^{-2} (i=1,
\cdots, 15)$ for models~II and III, respectively.
}
\label{fig:qxy}
\end{figure}

\begin{figure}
\includegraphics[keepaspectratio=true,width=8cm]{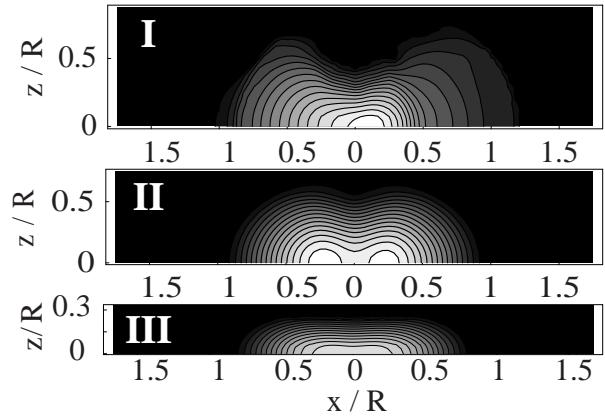}
\caption{
Density contours in the meridional plane for three differentially rotating
stars (see Table~\ref{tab:initial}).  The parameters and the contour
levels are the same as Fig.~\ref{fig:qxy}.
}
\label{fig:qxz}
\end{figure}

The density contour of the differentially rotating stars are shown in 
Fig.~\ref{fig:qxy} for the equatorial plane and in Fig.~\ref{fig:qxz}
for the meridional plane.  It is clearly seen in Fig.~\ref{fig:qxy}
that one-armed spiral structure is formed at the intermediate stage of
the evolution for model I, and that bar structure is formed for models
II and III once the dynamical instability sets in.

\begin{figure}
\includegraphics[keepaspectratio=true,width=8cm]{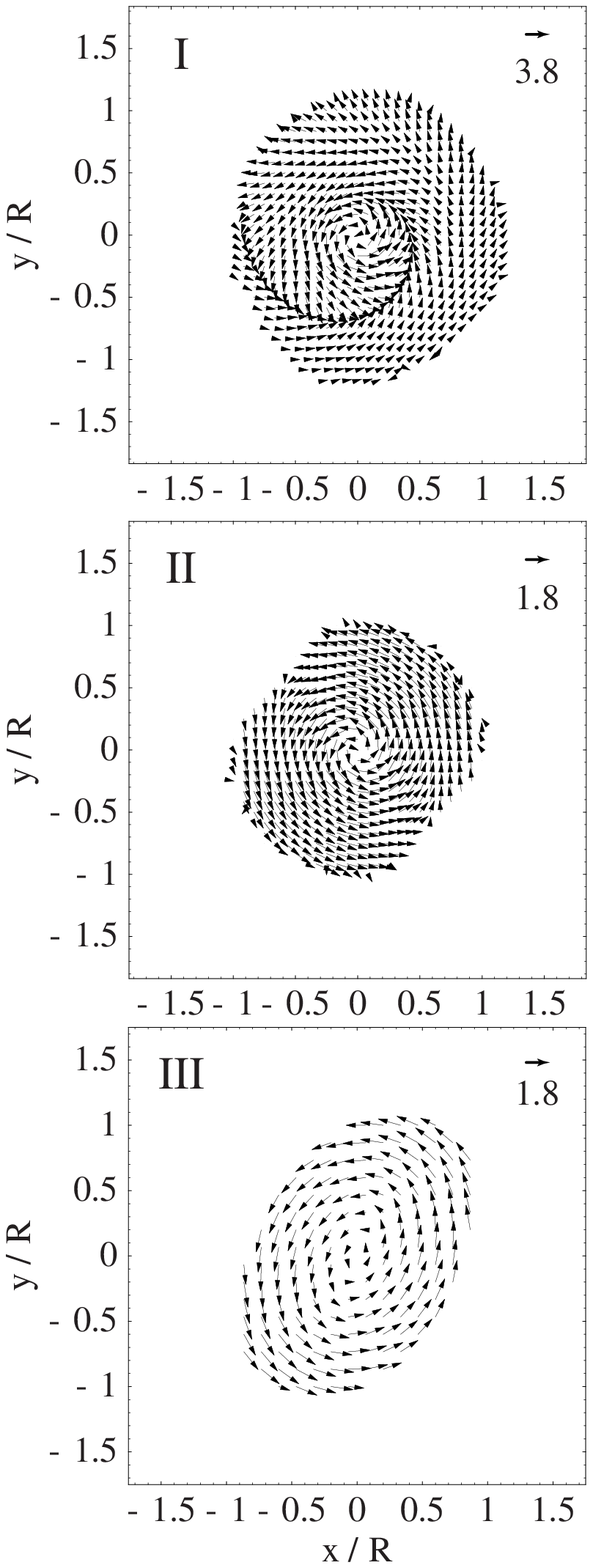}
\caption{
Velocity field ($v^{i} / |v_{\rm s}^{i~{\rm (0)}}|$) in the equatorial
plane for three differentially rotating stars (see
Table~\ref{tab:initial}).  The time for each snapshot is the same as
in Fig.~\ref{fig:qxy}.  Note that the velocity arrows are normalized
as indicated in the upper right hand corner of each snapshot.
$|v_{\rm s}^{i~{\rm (0)}}|$ denotes the surface absolute velocity at
$t=0$.
}
\label{fig:vxy}
\end{figure}

\begin{figure}
\includegraphics[keepaspectratio=true,width=8cm]{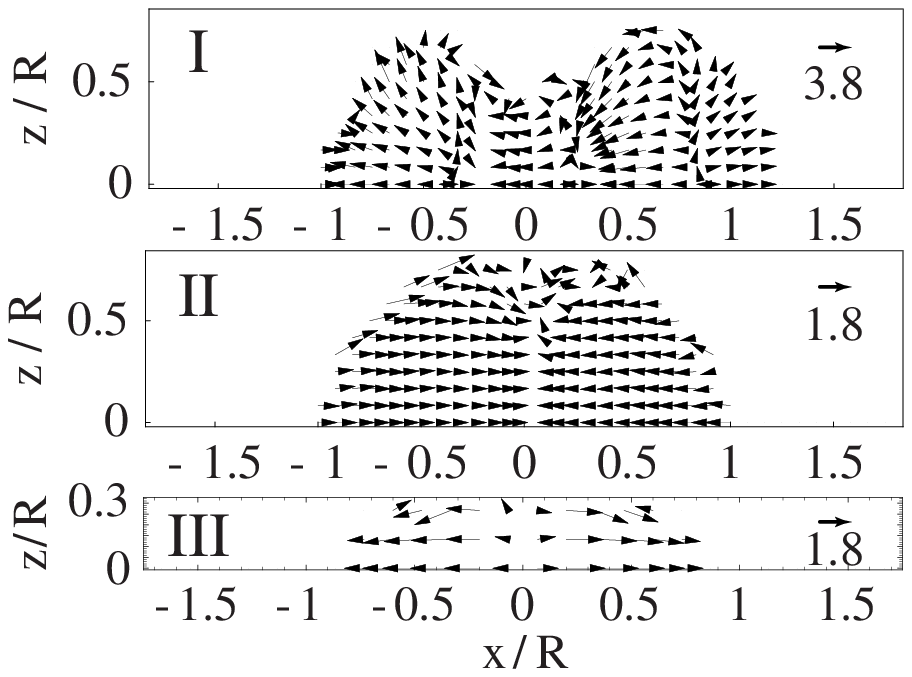}
\caption{
Velocity field ($v^{i} / |v_{\rm s}^{i~{\rm (0)}}|$) in the meridional
plane for three differential rotating stars (see Table~\ref{tab:initial}).
The time for each snapshot is the same as in Fig.~\ref{fig:qxy}.
}
\label{fig:vxz}
\end{figure}

We show velocity fields in Fig.~\ref{fig:vxy} in the equatorial plane
and in Fig.~\ref{fig:vxz} in the meridional plane during the
evolution.  Note that shocks occur during the formation of $m=1$
instability.  We also find that the fluid motion of the $z$-direction
does not play a dominant role in generating the dynamical instabilities.

\begin{figure}
\includegraphics[keepaspectratio=true,width=8cm]{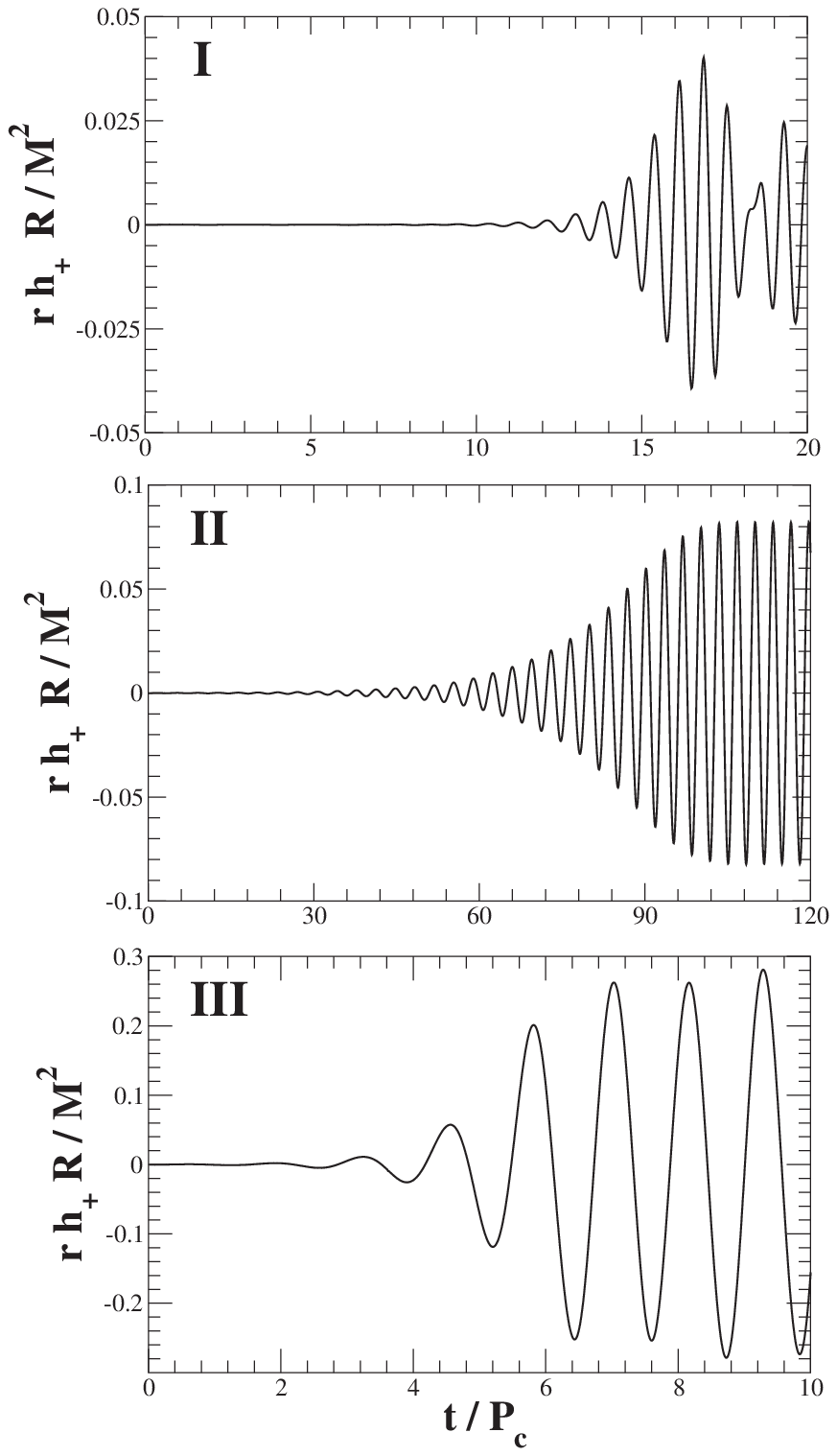}
\caption{
Gravitational waveform for three differentially rotating stars (see
Table~\ref{tab:initial}) as seen by a distant observer located on the
rotational axis of the equilibrium star.
}
\label{fig:gw}
\end{figure}

We also show gravitational waves generated from dynamical one-armed
spiral and bar instabilities in Fig.~\ref{fig:gw}.  For $m=1$ modes,
the gravitational radiation is emitted not by the primary mode itself,
but by the $m=2$ secondary harmonic which is simultaneously excited,
albeit at the lower amplitude.  Unlike the case for bar-unstable
stars, the gravitational wave signal does not persist for many
periods, but instead damp fairly rapidly.

\section{Stability analysis of a differentially rotating cylinder}
\label{sec:LinearAnalysis}
\subsection{Rotating selfgravitating cylinder model}
\citet{Ostriker65} numerically computed physical characters of
non-rotating infinite cylindrical masses.  We here follow his
treatment and introduce the normalization of variables.

A cylinder rotates around its axis with a given angular velocity
profile, which is a function of a cylindrical radial coordinate
$\varpi$.  An equilibrium of the cylinder is determined by the balance
between pressure gradient, centrifugal force and self-gravity of the
cylinder.  We also introduce an azimuthal angular coordinate
$\varphi$, and a $z-$coordinate set along the axis of the cylinder.

For a fluid equation of state, we assume to have a polytropic relation 
\begin{equation}
\rho = \rho_{\rm c} \bar{\theta}^N,\quad p = p_{\rm c}
\bar{\theta}^{N+1}, 
\end{equation}
where  $\rho_c$ and $p_c$ are the normalization factors for a mass
density and a pressure, which we choose those values on the rotational
axis of the cylinder, $N$ is a polytropic index.  A variable with a
bar denotes a normalized one.  As shall be seen below, it is possible
to construct a cylinder with a finite cylindrical radius by using
large $N$.  A non-rotating spheroid with $N\ge 5$ has an infinite
radius, while a cylinder with $N=25$ still has a finite cylindrical
radius.  This qualitative difference from the well-known characteristics 
of spheroid simply results from the difference of geometry.

We normalize the cylindrical radial coordinate as $\varpi = \alpha 
\bar{\varpi}$, where $\alpha = \sqrt{(N+1)/4\pi G}$, and the
rotational angular frequency as $\Omega = \bar{\Omega}\sqrt{4\pi
  G\rho_c}$.  Following a similar procedure to obtain Lane-Emden
equation of spherical polytropes, more specifically, with using
$\varpi-$ component of the Euler equation and Poisson equation for a
gravitational potential, we obtain Lane-Emden equation for a
differentially rotating cylinder as 
\begin{equation}
  \frac{d^2\bar{\theta}}{d\bar{\varpi}^2} + \frac{1}{\bar{\varpi}}
  \frac{d\bar{\theta}}{d\bar{\varpi}}
   + \bar{\theta}^N = 2\bar{\Omega}\frac{d}{d\bar{\varpi}}
\left[\bar{\varpi}\bar{\Omega}\right].
   \label{lane-emden}
\end{equation}
The rotational profile of an angular velocity we study here is the
same as the one used in the hydrodynamical simulations (equation~\ref{eqn:omega}),
\begin{equation}
  \bar{\Omega} = \frac{B}{\bar{\varpi}^2+A},
\end{equation}
where $A$ and $B$ are parameters.  This is the same as equation~(\ref{eqn:omega}), with $d=\sqrt{A}$ and $j_0 = B$.  For simplicity,
we hereafter omit `bars' from all the equations.

A frequently used dimensionless measure of rotation is $T/|W|$.  The 
rotational kinetic energy $T$ is defined as
\begin{equation}
  T = \frac{1}{2}\int\rho\varpi^2\Omega^2 dV,
\end{equation}
where integration is done for cylinder of unit length.  As to
gravitational energy, we follow the definition in \citet{CSS03} as 
\begin{equation}
W \equiv -\int\rho n^i\nabla_i\Phi dV
= -\frac{m(\varpi)^2}{4\pi}.
\end{equation}
Here, $\Phi$ is the gravitational potential and $n^i$ is a unit normal
vector of $\varpi=$const. surface.  The integration is performed for a
unit length along the axis.  $m(\varpi)$ is the mass contained inside
the cylindrical radius per unit length.

\begin{figure}
\includegraphics[keepaspectratio=true,width=8cm]{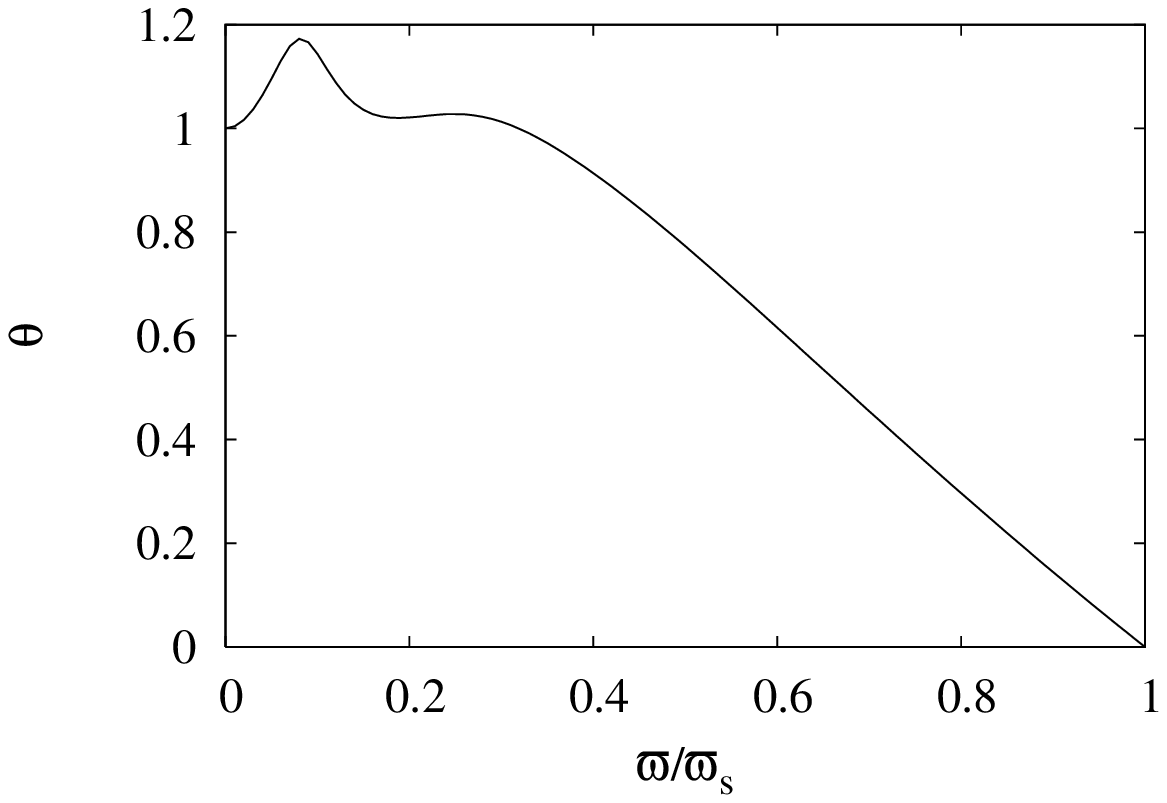}
\caption{Equilibrium profile of Lane-Emden function for a polytropic
  index $N=25$ and parameters of rotation $(A,B)=(0.43,1.6)$.  Note
  that $T/|W|=0.452$.  We find an $m=1$ unstable mode in corotation in
  this case. 
}
\label{cylinder-equil}
\end{figure}

For a large degree of differential rotation, the density maximum of
the configuration becomes off-centred.  An example of the profile of
Lane-Emden function in such a case is plotted in
Fig.~\ref{cylinder-equil}.

\subsection{Linear perturbation of a cylinder}
\subsubsection{Perturbation equations}
To study linearized oscillations of selfgravitating cylinder, we
simultaneously solve linearized version of (1) equation of continuity;
(2) Euler equation; (3) Poisson equation for a gravitational potential.
The adiabatic index of a perturbed fluid is assumed to coincide with
that of background one, so that no $g$-mode appears in our
computation.  We also assume that there is no motion along the
rotational axis of the cylinder, and therefore no dependence of the
quantities on $z$-coordinate.

Assuming a simple harmonic dependence of Eulerian perturbation of
variable $f$ as
\begin{equation}
\delta f (t,\varpi, \varphi) = 
\delta f_1(\varpi) \exp(-i\sigma t + im\varphi),
\end{equation}
we can write down the perturbed equations as follows.\\
Equation of continuity:
\begin{equation}
\frac{du}{d\varpi} + \frac{Ns}{\theta} q +
\left( \frac{N}{\theta} \frac{d\theta}{d\varpi} + 
\frac{1}{\varpi}\right) u -\frac{m v}{\varpi} = 0.
\label{continuity}
\end{equation}
$\varpi$-component of Euler equation:
\begin{equation}
s u + 2 \Omega v = \frac{dq}{d\varpi} + \frac{dy}{d\varpi}.
\label{euler-r}
\end{equation}
$\varphi$-component of Euler equation:
\begin{equation}
s~ v + \frac{\kappa^2}{2\Omega}u 
= \frac{m(q+y)}{\varpi},
\label{euler-phi}
\end{equation}
where $\kappa := \sqrt{2\Omega\left(2\Omega+\varpi
    d\Omega/d\varpi\right)}$ is the epicyclic frequency.\\
Poisson equation for a gravitational potential 
\begin{equation}
\frac{d^{2} y}{d \varpi^{2}} + \frac{1}{\varpi} \frac{dy}{d\varpi} - 
\frac{m^2}{\varpi^2} y = N \theta^{N-1} q.
\label{poisson}
\end{equation}
We have here defined Eulerian perturbation quantities as 
\begin{equation}
u = i \cdot \delta v_{1}^{\varpi}, \quad
v = \delta v_{1}^{\varphi}, \quad
y = \delta \Phi_{1}, \quad
q = \delta \theta_{1}.
\end{equation}
We also define the following quantity
\begin{equation}
s(\varpi) = \sigma - m \Omega(\varpi).
\end{equation}
If $s=0$ for a certain cylindrical radius, the equations becomes
singular there. 
\footnote{Unless the rotational angular frequency is pathological, we
  expect to have a regular singularity there.}  
We call it corotation singularity.  In that case, a corotation radius 
$\varpi_{\rm crt}$ is defined as this singular point.  Corotation
radius corresponds to a cylindrical surface on which the pattern speed
of the oscillation $\Re[\sigma/m]$ is equal to the local angular
frequency of the background flow, $\Omega(\varpi_{\rm crt})$.
Although $s=0$ is satisfied only for a purely real eigenfrequency, we
here denote that a mode is in corotation when a cylindrical radius
satisfies $\Re[\sigma]-m\Omega=0$ in the cylinder, even if we have a
complex eigenfrequency $\sigma$.  Thus a corotation radius is also
defined for complex mode where the equation has no singularity.

\begin{figure}
\includegraphics[keepaspectratio=true,width=8cm]{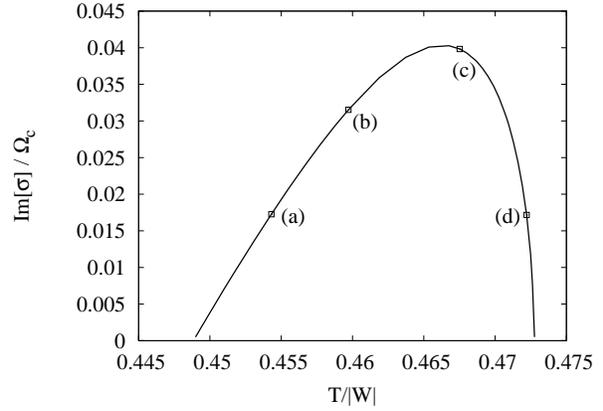}
\caption{Imaginary part of the eigenfrequency $\sigma$ for a
  dynamically unstable $m=1$ mode in a cylinder with $A=0.6$.  Note
  that a polytropic index $N$ is $25$.  The parameters of models
  marked as (a) to (d) are listed in Table~\ref{eqcyltab}.
}
\label{im-eigenfreq}
\end{figure}

To close the eigenvalue problem, we should impose boundary conditions.
One of the two conditions at the surface $\varpi=\varpi_{\rm s}$ where
equilibrium pressure becomes zero is the conventional free boundary
condition.  This means no stress is exerted on the cylindrical
surface, which reduces to the condition:
\begin{equation}
  sq + \frac{d\theta}{d\varpi} u = 0.
\end{equation}
The other is the condition on the perturbed gravitational potential.
From equation~(\ref{poisson}), the perturbed potential outside the
cylinder ($\theta=0$, $q=0$) is $y\sim \varpi^{\pm m}$.  At the
surface of the cylinder we impose the condition that the gravitational
potential smoothly matches to the non-diverging solution at infinity.
Therefore, the continuity condition of the potential requires
\begin{equation}
  \frac{dy}{d\varpi} + \frac{m}{\varpi} y = 0,
\end{equation}
at the cylindrical surface.  On the rotational axis of the cylinder, 
$\varpi=0$, we impose a regularity condition on the variables.

Our numerical method to solve this eigenvalue problem is
straightforward:  We use the conventional shooting method to find
eigenmodes.  The system of linearized equations are solved both from
the rotational axis and from the surface of the cylinder, with the
boundary conditions being taken into account.  At the intermediate
matching radius, we impose a condition that both solutions connect
smoothly.  This procedure picks up the physical solutions of the
eigenvalue problem. 

\begin{figure*}
\includegraphics[keepaspectratio=true,width=16cm]{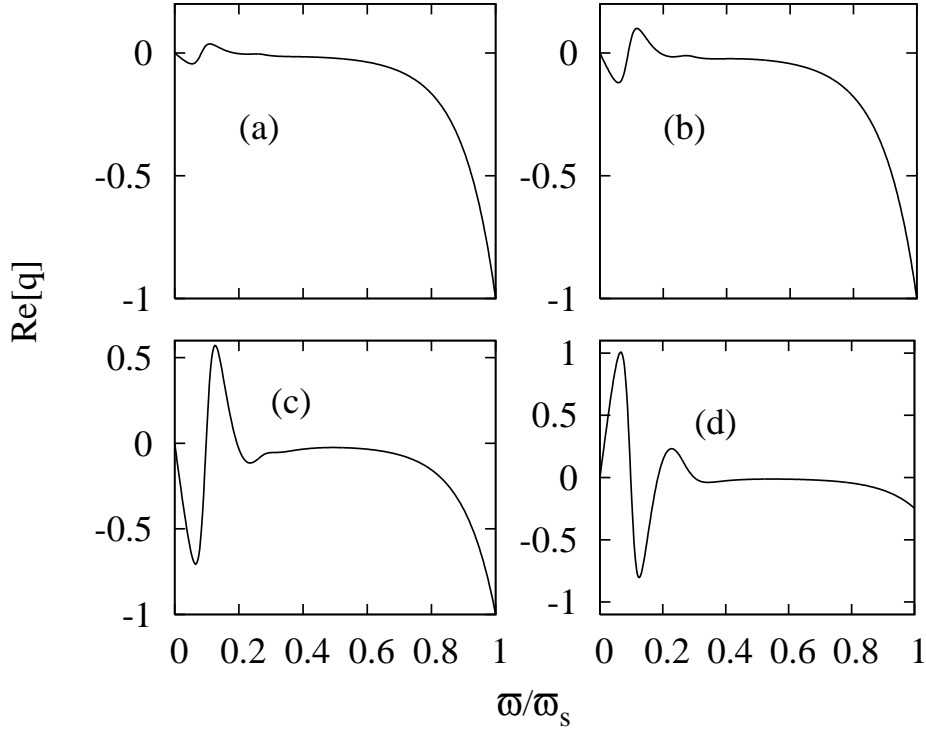}
\caption{
Eigenfunction of an unstable $m=1$ mode in $N=25$ cylinder.  We
normalize the cylindrical radius $\varpi$ with the surface radius
$\varpi_s$.  Real part of Lane-Emden function $q=\delta\theta$ are
plotted.  Note that the normalization of the eigenfunction is
arbitrarily.  The label (a) to (d) in the figure correspond to the
model in Table~\ref{eqcyltab}.
}
\label{deltatheta}
\end{figure*}

As we are interested in a dynamical instability, we assume the
eigenfrequency takes complex values as well as other perturbed
variables.  We focus to the case where $s(\varpi) = \sigma -
\Omega(\varpi)$ is non-zero, which is true except for a real $\sigma$
at the corotation region, i.e., $\Omega(\varpi = \varpi_{\rm s}) \le
\sigma \le \Omega(\varpi=0)$.  
\footnote{The pure corotation needs careful treatment around the
  singular point in order to pick up solutions as regular as possible
  We can use for instance Frobenius expansion \citep{RK01,WABS03}
  there.} 
As a result, our present code can compute modes without corotation 
singularity.  We also find convergence to `modes' whose frequency is
in the corotation region on the real axis of $\sigma$ plane.  Although
the `eigenfunction' of it suggest that we are picking up one of the
singular eigenmodes from continuous spectrum, we cannot prove it by
our present code.

\subsubsection{Characters of $m=1$ unstable eigenmode}
First we note that we find unstable $m=1$ mode only for extremely soft 
equation of state.  In general, we can construct a cylinder with a
finite cylindrical radius (but an infinite length for $z$-direction)
for an extremely large polytropic index $N$ than the case of a
spheroid.

From the results of hydrodynamical simulations \citep{SBS03}, we find
that an $m=1$ instability appears in a soft equation of state ($N \ga
2.5$).  However in case of a cylinder, $N$ should satisfy $N\ga 20$ to
find an $m=1$ instability.  This is a drawback of the cylinder to be
compared with a result of the spheroid, although the behaviour of the
unstable mode is similar to the spheroid.  We expect that a
qualitative nature of these modes are similar even the polytropic
indices are quite different.

For a fixed polytropic index, we have two parameters $A$ and $B$ to
specify an equilibrium model.  We construct a sequence fixing $A$ with
changing $B$, which roughly corresponds to changing $T/|W|$ in a fixed
degree of differential rotation.  We summarize the characteristics of
models studied in this paper in Table~\ref{eqcyltab}.

In Fig.~\ref{im-eigenfreq} we plot an imaginary part of the
eigenfrequency of $m=1$ mode as a function of $T/|W|$, fixing $A=0.6$
and $N=25$.  The mode has a corotation radius inside the cylinder.
The unstable mode appears only in a limited range of $T/|W|$.  This
behaviour of the imaginary part of the eigenfrequency is almost
insensitive to the degree of differential rotation which is
parametrized by $\sqrt{A}$.

\begin{table*}
\centering
\begin{minipage}{9.5cm}
\caption{Parameters for equilibrium models and eigenfrequency of $m=1$
  and $m=2$ modes.  Models (a)-(d) correspond to those in
  Fig.~\ref{deltatheta}.
\label{eqcyltab}}
\begin{tabular}{@{}ccccccc@{}}
\hline
Model & mode & $N$
&  $\Omega_{\rm c} / \Omega_{\rm s}$ 
& $T/|W|$
& ${\sigma/\Omega_c}^{a}$
& ${\varpi_{\rm crt}/\varpi_{\rm s}}^{b}$
\\
\hline
(a) & $m=1$ & 25 & 13.96 & 0.4543 & $0.5126+0.01726 i$ & 0.2709\\
(b) & $m=1$ & 25 & 11.34 & 0.4597 & $0.5512+0.03153 i$ & 0.2806\\
(c) & $m=1$ & 25 & 8.218 & 0.4675 & $0.6281+0.03982 i$ & 0.2864\\
(d) & $m=1$ & 25 & 6.067 & 0.4722 & $0.6910+0.01716 i$ & 0.2715\\
\hline
(b)-s1 & $m=1$ & 25 & 11.34 & 0.4597 & $-0.245~(\mbox{real})$ & ---\\
(b)-s2 & $m=1$ & 25 & 11.34 & 0.4597 & $1.15~(\mbox{real})$ & ---\\
\hline
(e) & $m=2$ &  1 & 13.00 & 0.170 & $0.3269 + 0.01256 i$ & 0.507\\
\hline
\end{tabular}

\medskip
${}^{a} \sigma$: eigenfrequency of the mode.
${}^{b} \varpi_{\rm crt}$: corotation radius; $\varpi_{\rm s}$: surface
  radius.
\end{minipage}
\end{table*}

Interestingly, this is reminiscent to the character of low-$T/|W|$ bar
($m=2$) instability which is recently found by \citet{SKE02,SKE03}
(see Fig.~3 in \citet{SKE03}).  For a fixed degree of differential
rotation, $T/|W|$ that permits unstable mode is limited in a finite
range.  This may suggest that $m=1$ instability and low-$T/|W|$ bar
instability may have the same (or similar) generation mechanism.

We note that a real part of eigenfrequency is monotonically increasing 
function of $T/|W|$ and of the degree of differential rotation (see
Table~\ref{eqcyltab}).  The frequency is the order of unity for all cases
here.  For each equilibrium model with a sufficiently large degree of
differential rotation and a limited range of $T/|W|$, we find unstable
$m=1$ modes in the corotation.  Note that we also have exponentially
damping modes, whose eigenfrequency are the complex conjugate of the 
dynamically unstable modes.  We also find discrete stable modes
outside the corotation region, which have a purely real
eigenfrequency.  For equilibrium models where we found unstable $m=1$ 
corotating modes, we did not find any dynamically unstable $m=1$ mode
outside the corotation.

\begin{figure}
\includegraphics[keepaspectratio=true,width=8cm]{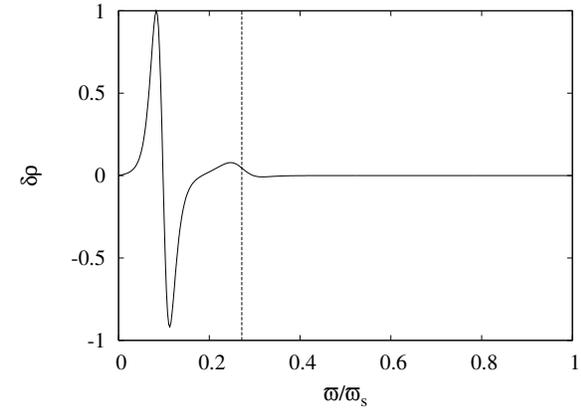}
\caption{
Perturbed mass density of an unstable $m=1$ mode for model (b) (see
Table~\ref{eqcyltab}).  As it is a solution of linear problem, the scaling is
arbitrarily chosen. 
}
\label{fig:ptb_cylinder}
\end{figure}

\begin{figure}
\includegraphics[keepaspectratio=true,width=8cm]{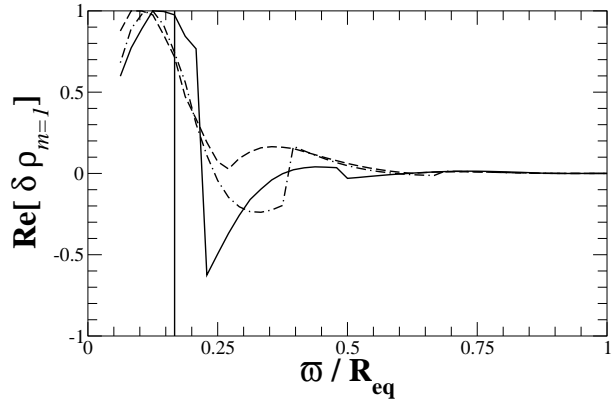}
\caption{
Amplitude of perturbed mass density of an unstable $m=1$ mode for 
model I (See Table~\ref{tab:initial}).  Note that the amplitude is
normalized with its maximum.  Solid, dashed, and dash-dotted lines
represent $t = 8.15 P_c$, $9.32 P_c$, and $10.48 P_c$, respectively.
Vertical line represents the corotation radius of the rotating star.
}
\label{fig:ptb_star}
\end{figure}

We show the real part of the eigenfunction $q$ of an $m=1$ unstable
mode in Fig.~\ref{deltatheta}. In order to compare the $m=1$
eigenfunction of the linear analysis with that of the hydrodynamical
simulation, we plot a perturbed mass density in a cylinder model 
(Fig.~\ref{fig:ptb_cylinder}) and a perturbed $m=1$ unstable mass
density in a differentially rotating star (Fig.~\ref{fig:ptb_star}).
In order to compute the perturbed $m=1$ unstable mass density in a
differentially rotating star, we follow
\begin{eqnarray}
\delta \rho &=& \rho(t) - \rho_{\rm eq}
, \\
\delta \rho_{m=1} &=&
\frac{1}{2 \pi} \int \delta \rho~e^{-i\varphi} d \varphi
.
\end{eqnarray}
We have a common feature that the $m=1$ unstable mass density has a
single oscillation inside the star and in the cylinder.  However the
behavior is quite different:  for a cylinder the oscillation is
concentrated inside the corotation radius, while for the
differentially rotating star the oscillation is spread out to the
whole equatorial surface radius.

\begin{figure}
\centering
\includegraphics[keepaspectratio=true,width=9.0cm]{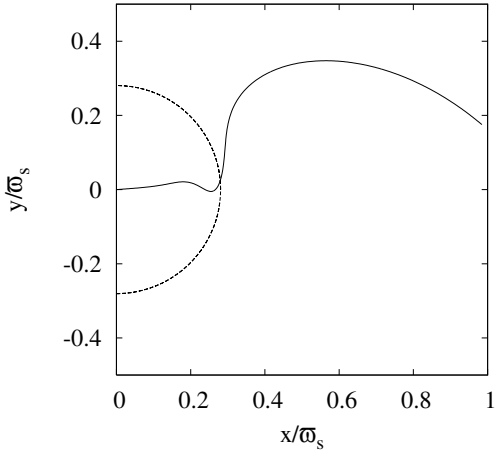}
\caption{
Phase-constant curve of velocity perturbation $\delta v^{\varpi}$ for
model (b) (see Table~\ref{eqcyltab}). The curve is plotted in the plane
perpendicular to the rotational axis. The radius of cylinder is
normalized to be unity.  The dotted half-circle marks the corotation
radius of the mode.  Note that the direction of rotation of background
flow is clockwise.
}
\label{fig:phase_cylinder}
\end{figure}

\begin{figure}
\centering
\includegraphics[keepaspectratio=true,width=6.5cm]{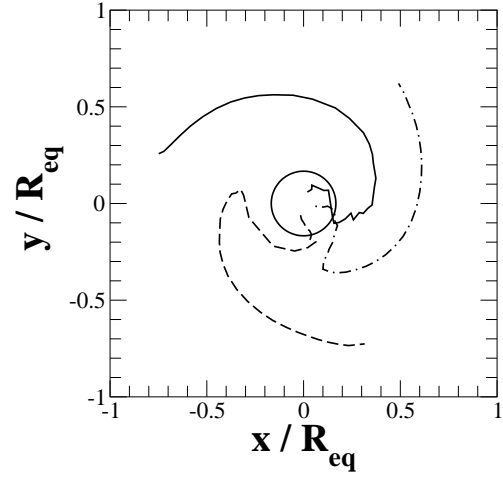}
\caption{
Phase-constant curve of perturbed radial velocity in the equatorial
plane of an unstable $m=1$ mode for model I (see
Table~\ref{tab:initial}).  The snapshot times are the same as that in
Fig.~\ref{fig:ptb_star}.  Solid circle denotes the corotation radius
of the star.  Note that the direction of rotation of background flow
is counter-clockwise.
}
\label{fig:phase_star}
\end{figure}

In order to focus on the comparison of the spiral structure of the
$m=1$ unstable mode, we introduce a phase-constant curve of a
perturbed radial velocity.  In the linear analysis, complex velocity
$\delta v^{\varpi}$ is written as (we omit the factor of time
dependence $e^{-i\sigma t}$ since it is not relevant to a momentary
spacial pattern)
\begin{equation}
\delta v^{\varpi} = U(\varpi) e^{iS(\varpi)+im\varphi},
\end{equation}
where $U(\varpi)$ is a real amplitude and $S(\varpi)$ is a phase
function.  An equation defining phase-constant pattern is therefore,
\begin{equation}
S(\varpi)+m\varphi = \tan^{-1}\frac{\Im[\delta
v^{\varpi}]}{\Re[\delta v^{\varpi}]} + m\varphi \equiv {\rm
const.} 
\end{equation}
In a similar way, we also obtain a phase-constant pattern from the
result of nonlinear hydrodynamical simulation.  First we expand a
perturbed radial velocity in the equatorial plane in terms of
$\varphi$ as
\begin{equation}
\delta v^{\varpi} = \delta v^{\varpi}_{m} e^{i m \varphi}
  = \| \delta v^{\varpi}_m \| ~ e^{i (m \varphi + S_m)}
,
\end{equation}
where
\begin{eqnarray}
\delta v^{\varpi}_{m} &=&
  \frac{1}{2 \pi} \int \delta v^{\varpi} ~e^{-i m \varphi} d \varphi
,\\
\| v^{\varpi}_m \| &=& 
  \sqrt{(\Re [\delta v^{\varpi}_m])^2 + (\Im [\delta v^{\varpi}_m])^2}
,\\
S_m &=& \tan^{-1} \frac{\Im [\delta v^{\varpi}_m]}{\Re [\delta v^{\varpi}_m]}
.
\end{eqnarray}
Next, we focus on the phase-constant curve $m\varphi + S_m = C$, where 
$C$ is a constant.  Hereafter we choose $C=0$ since it only shifts the
azimuthal angle.  The $m=1$ phase constant curve of the perturbed
radial velocity in the equatorial plane is 
\begin{eqnarray}
x &=& \varpi \cos \varphi 
  = \frac{\varpi}{\| \delta v^{\varpi}_1 \|} 
  \Re [\delta v^{\varpi}_1]
, \\
y &=& \varpi \sin \varphi = 
  - \frac{\varpi}{\| \delta v^{\varpi}_1 \|} 
  \Im [\delta v^{\varpi}_1]
.
\end{eqnarray}
We compare the phase-constant curve of $m=1$ unstable mode between
a cylinder (Fig.~\ref{fig:phase_cylinder}) and a star
(Fig.~\ref{fig:phase_star}).  Both figures clearly have an $m=1$
unstable spiral structure.  However the direction of arms are
opposite.  Also trailing or leading nature of arms depends on the
radial distance from the center in both cases.  For the cylinder model
in Fig.~\ref{fig:phase_cylinder}, the arm changes from leading to
trailing if we follow it from the center. For the star in
Fig.~\ref{fig:phase_star}, an arm is leading inside while trailing
outside.  This apparent difference, however, does not prevent us from
comparing these two models. In fact, according to the result by 
\citet{Robe79} leading/trailing nature of spiral arm is not relevant to
stability nature nor classification of eigenmodes.  The same unstable
eigenmode can have an arm with trailing, leading or mixed direction,
depending on the equilibrium parameter. Thus the apparent difference
is not significant here. The important point here is that the two
model share one-armed spiral characteristics.

\section{Canonical Angular Momentum to Diagnose Dynamical Instability}
\label{sec:Canonical}
\subsection{Introduction of canonical angular momentum}
In order to diagnose the oscillations in a rotating fluid, we here
introduce the canonical angular momentum following \citet{FS78a}.  In
the theory of adiabatic linear perturbations of a perfect fluid
configuration with some symmetries, it is possible to introduce
canonical conserved quantities associated with the symmetries.  When
we introduce Lagrangian displacement vector $\xi^i$, which is a vector
connecting a perturbed fluid element to a corresponding non-perturbed
one, the linearized perturbation equations are cast into a single
second order differential equation for $\xi^i$
\citep{FS78a};
\begin{equation}
A^i_j \partial_t^2\xi^j + B^i_j \partial_t\xi^j + C^i_j \xi^j = 0.
\label{eq_motion}
\end{equation}
Here the operators $A$ and $C$ are Hermitian, $B$ anti-Hermitian,
respectively up to divergence term [see equations~(32) to (34) in
\citet{FS78a} for the precise expressions], with respect to an inner
product of displacements $\eta^i$ and $\xi^i$,
\begin{equation}
<\eta,\xi> \equiv \int {\eta_i}^*\xi^i dV
	 = \int {\eta^i}^*\xi_i dV,
\end{equation}
where $\eta^*$ means Hermite conjugate of $\eta$.

The master equation~(\ref{eq_motion}) is derived from the variational 
principle with an action
\begin{equation}
I = \int_{t_1}^{t_2}dt \int dV~ {\cal L},
\end{equation}
where Lagrangian density ${\cal L}$ is defined by
\begin{equation}
{\cal L} := \frac{1}{2}
\left[
  (\partial_{t}{\xi^*_i}) A^i_j (\partial_{t}{\xi}^j) +
  (\partial_{t}{\xi^*_i}) B^i_j \xi^j - \xi^*_i C^i_j \xi^j
\right].
\end{equation}

Applying N\"{o}ther's theorem to this Lagrangian, we obtain canonical 
conserved quantities \citep{Wald84}.  In particular, if we have an
axisymmetry in the background fluid, the corresponding Killing vector
$\partial_\varphi$ produces a conserved current related to the angular
momentum,
\begin{equation}
{\cal J}^\alpha = \partial_{_\varphi}^\alpha {\cal L}-
\frac{\partial {\cal L}}{\partial(\partial_\alpha\xi^j)}
\pounds_{_{\partial_{_\varphi}}}\xi^j -
\frac{\partial {\cal L}}{\partial(\partial_\alpha\xi^{j*})}
\pounds_{_{\partial_{_\varphi}}}\xi^{j*},
\label{angularmomentum-current}
\end{equation}
where $\partial_\varphi^\alpha$ is $\alpha-$component of vector 
$\partial_\varphi$, and $\pounds_{v}$ denotes Lie derivative along a
vector $v$.  Note that our variable $\xi^i$ is a complex vector field.
The time component of this current is a canonical angular momentum
density, while spacelike components are the flux density.  In the rest
of this section, let us follow \citet{FS78a} to derive an explicit
form of the canonical angular momentum.

A natural symplectic structure is introduced as an inner product of a
field $\xi$ and its conjugate momentum density,
\begin{equation}
W(\eta,\xi) := 
\left< 
  \eta, A (\partial_{t}{\xi}) + \frac{1}{2} B \xi 
\right>
 - 
\left< 
  A (\partial_{t}{\eta}) + \frac{1}{2} B \eta, \xi 
\right>.
\end{equation}
It is easy to find that this product is conserved using the master
equation~(\ref{eq_motion}) and using the symmetric property of
operators $A$, $B$, $C$ as 
\begin{equation}
\partial_t W(\eta,\xi) = 0,
\end{equation}
where both $\eta^i$ and $\xi^i$ are solutions to the master equation~(\ref{eq_motion}).  We obtain the canonical angular momentum of the
system when the background fluid is axisymmetric (thus $A,B,C$ commute
with $\partial_\varphi$),
\begin{equation}
   J_c(\xi) = -\frac{1}{2}W(\partial_\varphi\xi,\xi),
\end{equation}
which is a volume integral of the canonical angular momentum density
defined by equation~(\ref{angularmomentum-current}).  From the
definition, these conserved quantities are quadratic in the Lagrangian
displacement vector.

These quantities, however, are `gauge-dependent' in general.  This
means that we may find `trivial displacement' vector $\zeta^i$ for any
physical solution of the master equation~(\ref{eq_motion}).  The
trivial displacement is added to `physical' solution $\xi^i$ to
produce a different displacement, which corresponds to the same
physical solution (i.e., it does not change Eulerian perturbation of
physical quantities).  The trivial defines a class of gauge
transformation of the same physical solution, under which the
canonical energy or momentum are generally not invariant.  Thus a
naive use of it to the stability problem of a fluid may lead to a
wrong conclusion.  \citet{FS78a} showed that we can find a class of a
physical solution to the master equation~(\ref{eq_motion}) called
`canonical displacement' orthogonal to the trivials, for which we have
no contribution from the trivial displacement to the canonical
quantities.  Fortunately, in the case we are interested in
(normal-mode problem with non-zero complex frequency), the
displacement are always orthogonal to the trivials \citep{FS78b}.

Finally, we remark that the canonical energy or angular momentum of
the dynamically unstable modes are zero.  This is simply found if we
are conscious of the existence of an imaginary part $\sigma_I$ in the 
eigenfrequency $\sigma$.  Then the conservation equation of the
product $W$ is written as 
\begin{equation}
\partial_t W = 2\sigma_I W=0,
\end{equation}
and therefore $W=0$.

\begin{figure*}
\centering
\includegraphics[keepaspectratio=true,width=16cm]{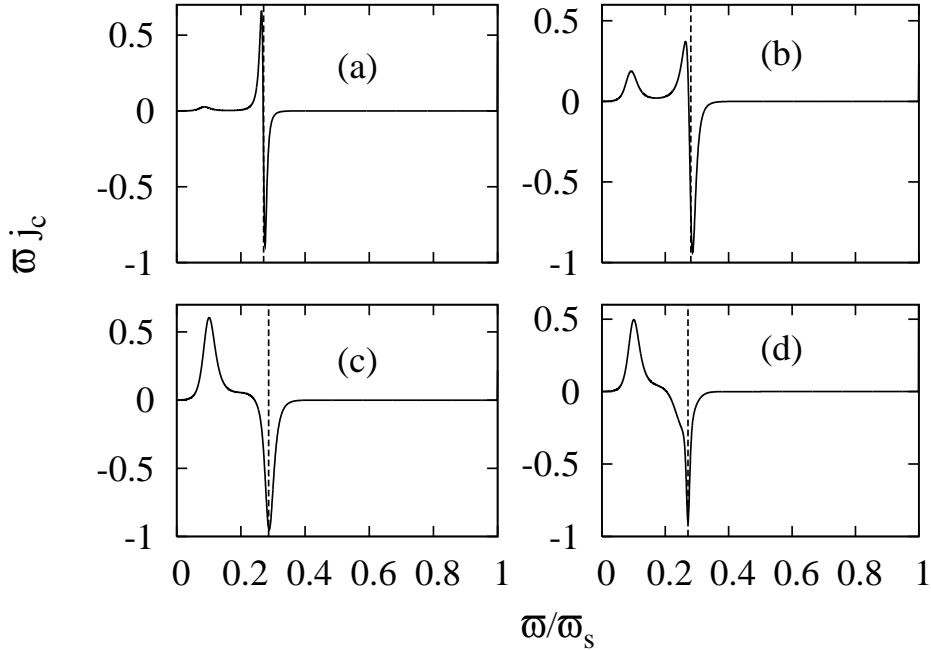}
\caption{
Distribution of the canonical angular momentum density for $m=1$
unstable mode.  Plots are integrand of equation~(\ref{canonJform}),
$\varpi j_c (\varpi)$.  Models (a) to (d) are the same as in
Fig.~\ref{deltatheta}.  Vertical dashed lines mark the corotation
radius of the mode.
}
\label{jc-distrib-m1-unstab}
\end{figure*}

Next we write down the explicit form of canonical angular momentum
used in the following discussion.
\begin{eqnarray}
 J_{\rm c}(\xi) &=& -\frac{1}{2}\hat{W}(\partial_\varphi\xi,\xi)
\nonumber \\
&=&
- \frac{1}{2} 
\left< 
  \partial_\varphi \xi, A (\partial_t \xi) + \frac{1}{2} B \xi 
\right>
\nonumber \\
&&
+ \frac{1}{2} 
\left< 
  A \partial_\varphi \partial_t \xi + \frac{1}{2} B \xi, \xi 
\right>.
\end{eqnarray}
Thus,
\begin{eqnarray}
 J_{\rm c}(\xi) &=&
 -\frac{1}{2} \int_{V} \partial_\varphi \xi_i^*A^i_j \partial_t \xi^j
   dV 
 -\frac{1}{4}\int_{V}\partial_\varphi\xi_i^*B^i_j \partial_t \xi^j dV
\nonumber\\
&& 
 +\frac{1}{2}\int_{V}A^i_j \partial_t \partial_\varphi\xi^{j*}
   \cdot\xi_i dV 
 +\frac{1}{4}\int_{V}B^i_j\partial_\varphi\xi^{j*}\cdot\xi_i dV.
\nonumber \\
\label{def_J}
\end{eqnarray}
Note that since $A$, $B$, $C$ are defined in the background fluid
whose physical quantities are purely real, they are `real' quantities
(although $B$ is anti-Hermite as an operator).  As we are interested
in a normal mode solution with harmonic dependence in $t$ and
$\varphi$, the displacement vector can be written as 
\begin{equation}
\xi   = \xi_{0}(\varpi) e^{-i\sigma t+im\varphi},\quad
\xi^* = \xi^{*}_{0}(\varpi) e^{i\sigma^* t-im\varphi}.
\end{equation}
The first and third terms in equation~(\ref{def_J}) are combined to
produce
\begin{equation}
m\Re[\sigma]\int_{V} \rho|\xi|^2 dV,
\end{equation}
while the second and fourth are simplified as
\begin{eqnarray}
&&\frac{im}{4}\int_{V}
\left[
  \rho(\xi_i^*v^k\nabla_k\xi^i-\xi_iv^k\nabla_k\xi^{i*})
  + \nabla_j(\rho v^j \xi_k^*\xi^k)
\right]
\nonumber\\
&&-\frac{im}{4}\int_{V}
\left[
  \rho(\xi_iv^k\nabla_k\xi^{i*}-\xi_i^*v^k\nabla_k\xi^i)
  + \nabla_j(\rho v^j \xi_k^*\xi^k)
\right]
\nonumber\\
&=& -m\int_{V}\rho\cdot
\Im\left[\xi_i^*v^k\nabla_k\xi^i\right]dV
.
\end{eqnarray}
Note that $B$ is antisymmetric up to a divergence term which appears
in the integral above.  We have an exact cancellation to the
contribution of $B$.  As we are interested in the case with a circular
flow as a background whose nonzero component of velocity is
$v^\varphi=\Omega(\varpi)$, we can easily find that
\begin{equation}
\xi_i^*v^k\nabla_k\xi^i = 
im\Omega|\xi|^2 - \varpi\Omega\xi^{\varpi*}\xi^{\varphi}
+ \varpi\Omega\xi^{\varphi *}\xi^\varpi.
\end{equation}
Finally, we get the simple form of the canonical angular momentum
$J_{\rm c}$ as \footnote{\citet{NGG87} and \citet{CN92} derived the
  same formula to study the oscillations of the slender annuli around
  a point mass gravity source.  Watts (private communication) derived
  a formula of the canonical energy of eigenmodes in differentially
  rotating shell of fluid.}
\begin{equation}
J_{\rm c} = 
  m\int_{V}(\Re[\sigma]-m\Omega)\rho|\xi|^2 dV
 -2m\int_{V} \rho \varpi\Omega\cdot \Im[\xi^\varpi\xi^{\varphi *}] dV.
\label{canonJform}
\end{equation}

\subsection{Application to oscillations of a cylinder}
\label{subsec:AppCylinder}
We here present typical distribution of the canonical angular momentum
in the cylinder model.  The absolute amplitude of the plotted function
here has no significance, since a linear eigenfunction can be scaled
arbitrarily.

\begin{figure}
\centering
\includegraphics[keepaspectratio=true,width=8cm]{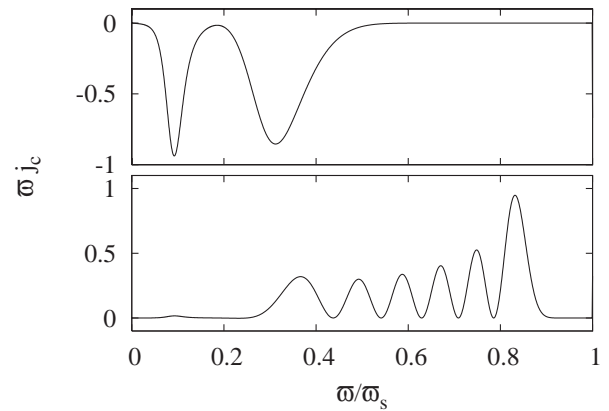}
\caption{
Distribution of the canonical angular momentum density for $m=1$
stable modes.  Plots are integrand of equation~(\ref{canonJform}),
$\varpi j_c(\varpi)$.  The parameters of top and bottom panels
correspond to (b)-s1 and (b)-s2 in Table~\ref{eqcyltab}.  The model
(b)-s1 satisfies $\sigma-\Omega(\varpi)<0$, while the model (b)-s2
satisfies $\sigma-\Omega(\varpi)>0$ throughout the fluid.
}
\label{jc-distrib-m1-stab}
\end{figure}

In Fig.~\ref{jc-distrib-m1-unstab}, we plot the integrand of canonical
angular momentum, defined from equation~(\ref{canonJform}) as
\begin{equation}
\varpi j_{\rm c}(\varpi) = m (\Re[\sigma]-m\Omega)\rho|\xi|^2
  - 2 m \rho \varpi\Omega\cdot \Im[\xi^\varpi\xi^{\varphi *}],
\end{equation}
for unstable $m=1$ mode.  Here, $j_c$ is the canonical angular momentum 
density.  Note that an integral in the entire cylinder is zero for
these cases.  The features of the canonical angular momentum
distribution for $m=1$ unstable modes are, 
\begin{enumerate}
\item 
It changes sign around corotation radius $\varpi_{\rm crt}$.
\label{ite:crt}
\item
It is positive for $\varpi<\varpi_{\rm crt}$, while 
negative for $\varpi>\varpi_{\rm crt}$. 
\label{ite:pnd}
\end{enumerate}
The feature~\ref{ite:crt} is remarkable and suggests us that the
instability is related to the corotation.  The feature has a clear
contrast for a stable mode (Fig.~\ref{jc-distrib-m1-stab}).  The
canonical angular momentum is either positive or negative definite,
and it does not change its sign.  Note that the former is the case
when the pattern speed of mode is faster than the rotation of cylinder
everywhere, while the latter is the opposite.  This feature is
expected from the equation~(\ref{canonJform}), if the first term is
dominant.  In such case, the sign of $\Re[\sigma]-m\Omega(\varpi)$
determines the sign of the canonical angular momentum.  This simple
interpretation, however, does not hold for the dynamically unstable
mode.  As it is shown in the feature~\ref{ite:pnd} above, we have a
positive canonical angular momentum inside the corotation, which is
opposite to the sign of $\sigma-\Omega(\varpi)$ for
$0\le\varpi<\varpi_{\rm crt}$.

\begin{figure}
\centering
\includegraphics[keepaspectratio=true,width=8cm]{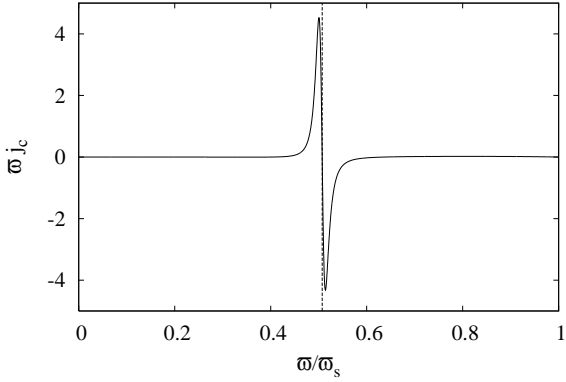}
\caption{
Distribution of canonical angular momentum density for $m=2$ unstable
modes (model~(e) in Table~\ref{eqcyltab}).  Plots are integrand of
equation~(\ref{canonJform}), $\varpi j_c(\varpi)$.  The vertical
dashed line marks the corotation radius.
}
\label{jc-distrib-m2-unstab}
\end{figure}

In Fig.~\ref{jc-distrib-m2-unstab}, we show an example of canonical
angular momentum distribution for $m=2$ unstable mode of
differentially rotating cylinder, which may be compared with the low
$T/|W|$ bar instability of \citet{SKE02,SKE03}.  We did not find $m=2$
unstable modes for the same parameters as in the case of $m=1$
instability.  The features at the corotation radius, however, are the
same as in $m=1$ instability.

It is interesting to see how the profile of the canonical angular
momentum changes when we consider the classical bar instability with
uniform rotation.  Unfortunately the bar mode of uniformly rotating
cylinder has a neutral stability point at the breakup limit
\citep{Luyten90}.  We instead looked at $m=2$ instability of uniformly
rotating, incompressible Bardeen disk \citep{Bardeen75} and the
classical bar instability of Maclaurin spheroid (see Appendix
\ref{appendix-classical} for these computation).  These are actually
more suitable for comparison to differentially rotating spheroidal
model, which we present in the following section.  For both of the
models we have analytic expressions of oscillation modes [\citet{SB83}
for a Bardeen disk and \citet{Chandra69} for Maclaurin spheroid].  It
is remarkable that the canonical angular momentum density is zero
everywhere (which ensures that the total canonical angular momentum
vanishes).  This is in a clear contrast with the $m=2$ instability in
the cylinder with highly differential rotation.

\subsection{Differentially rotating star}
We here present the application of the canonical angular momentum to
our hydrodynamics results.  Following three assumptions are made to
adopt our hydrodynamic results of dynamical instability (for both
$m=1$ and $m=2$) to the perturbative approach;
\begin{enumerate}
\item All physical quantities is in coherent oscillation of growing
  mode. 
\label{ite:same}
\item For each model, a growing mode with single $m$ is dominant.
\label{ite:dominant}
\item The motion of $z$-direction does not contribute to the
  instability. 
\label{ite:NoZ}
\end{enumerate}

From assumption~\ref{ite:same} we introduce a complex frequency
$\sigma$ that represents the same growing mode $m$.  Therefore all
physical quantities $f(t)$ that have a time dependence should satisfy 
\begin{equation}
f(t) = f_{1} \exp(- i \sigma t),
\end{equation}
where $\sigma = \sigma_{\rm R} + i \sigma_{\rm I}$, $f_{1}$ is a
complex quantity.  Note that $\sigma_{\rm R}$ and $\sigma_{\rm I}$ are
real quantities.

The assumption~\ref{ite:dominant} comes from the fact that single $m$ 
mode has a dominant contribution to the dynamical instability in the 
diagnostics (Fig.~\ref{fig:dig}).  Therefore all physical quantity 
$f(\varpi, \varphi)$ that have a dependence of azimuthal angle should
satisfy 
\begin{equation}
f(\varpi, \varphi) = f(\varpi) \exp(i m \varphi),
\end{equation}
where
\begin{equation}
f(\varpi) = \frac{1}{2\pi} \int_{0}^{2 \pi} d \varphi ~ 
             f(\varpi, \varphi) \exp(- i m \varphi).
\end{equation}

From the velocity snapshots in the meridional plane
(Fig.~\ref{fig:vxz}), the motion of the fluid along the rotation axis
does not have a significant contribution to the instability.
Therefore we can neglect the $z$-dependence in the function
(assumption~\ref{ite:NoZ}) as
\begin{equation}
f(\varpi, z) = f(\varpi).
\end{equation}

From the three assumptions, we determine the frequency from the dipole 
and quadrupole diagnostics as
\begin{eqnarray}
D(t) &=& \frac{1}{M} \int dv \rho(t, \varpi, \varphi, z) 
         \exp(i \varphi) 
\nonumber \\
&=&  \exp(-i \sigma t) \frac{1}{M} \int d\varpi
     2 \pi \varpi  \rho(\varpi),
\label{eqn:ddig}
\\
Q(t) &=& \frac{1}{M} \int dv \rho(t, \varpi, \varphi, z) 
         \exp(2 i \varphi)
\nonumber \\
&=&  \exp(-i \sigma t) \frac{1}{M} \int d\varpi
2 \pi \varpi  \rho(\varpi).
\label{eqn:qdig}
\end{eqnarray}

With the formulae above, we extract the complex frequency by fitting 
evolutions of dominant diagnostics for each cases.

Note that we averaged the frequency in the early stage of the
evolution, since the real part of the frequency is almost the same.
Note also that $D(0) \not = 0$ and $Q(0) \not = 0$ since we put $m=1$
and $m=2$ density perturbation at $t=0$ (equation~\ref{eqn:DPerturb}) to initiate $m=1$ or $m=2$ dynamical instability.

\begin{table}
\centering
\caption
{Eigenfrequency and the corotation radius of three differentially rotating
  stars 
\label{tbl:corfreq}}
\begin{tabular}{@{}ccc@{}}
\hline
Model &
$\sigma$ $[\Omega_{\rm c}]$ &
$\varpi_{\rm crt}$  $[R_{\rm eq}]$
\\
\hline
I & $0.590 + 0.0896 i$ & $0.167$
\\
II & $0.284 + 0.0121 i$ & $0.492$ 
\\
III & $0.757 + 0.200 i$ & ---
\\
\hline
\end{tabular}
\end{table}

We summarize the corotation radius and the complex frequency in
Table~\ref{tbl:corfreq} from Fig.~\ref{fig:dig}.  Eulerian perturbed
velocity is defined as
\begin{equation}
\partial_t v^{i} (t, \varpi, \varphi) = 
v^{i} (t, \varpi, \varphi) - v^{i}_{\rm eq} (\varpi),
\end{equation}
where $v^{i} (t, \varpi, \varphi)$ is the velocity at $t$, $v^{i}_{\rm
  eq} (\varpi)$ is the velocity at equilibrium, respectively.
Following assumption~\ref{ite:dominant}, we find,
\begin{equation}
\partial_t v^{i} (t, \varpi) =
\frac{1}{2\pi} \int_{0}^{2 \pi} d\varphi ~
\partial_t v^{i} (t, \varpi, \varphi) \exp(-i m \varphi).
\end{equation}

The Lagrangian displacement $\xi^{i}$ should satisfy the following
differential equation: 
\begin{equation}
\partial_{t} \xi^{i} = 
\partial_t v^{i} + \xi^{k} \partial_{k} v^{i}_{\rm eq} - 
v^{k}_{\rm eq}  \partial_{k} \xi^{i}.
\end{equation}
Using the three assumptions, $\varpi$ and $\varphi$ component of the 
Lagrangian displacement should be written as 
\begin{eqnarray}
\xi^{\varpi} &=& 
\frac{i \partial_t v^{\varpi}}{\sigma - m \Omega_{\rm eq}}
,\\
\xi^{\varphi} &=&
\frac{i \partial_t v^{\varphi}}{\sigma - m \Omega_{\rm eq}}
- \frac{(\partial_t v^{\varpi}) (\partial_{\varpi} \Omega_{\rm eq})}
{(\sigma - m \Omega_{\rm eq})^{2}}
,
\end{eqnarray}
where $\Omega_{\rm eq}$ denotes the angular velocity at equilibrium.

Using the Lagrangian displacement extracted by the formula above,
we compute the canonical angular momentum density defined as an
integrand of the canonical angular momentum in
equation~(\ref{canonJform}) as 
\begin{equation}
J_{\rm c} = \int d\varpi~ \varpi j_{\rm c}(\varpi),
\end{equation}
where
\begin{equation}
\varpi j_{\rm c}(\varpi) =  
  m (\Re[\sigma] - m \Omega) \rho|\xi|^2
 - 2 m \rho \varpi \Omega \cdot \Im[\xi^\varpi \xi^{\varphi *}]
.
\end{equation}
Note that we have used the assumptions~\ref{ite:same} -- \ref{ite:NoZ}
to compute the canonical angular momentum density.

\begin{figure}
\centering
\includegraphics[keepaspectratio=true,width=8cm]{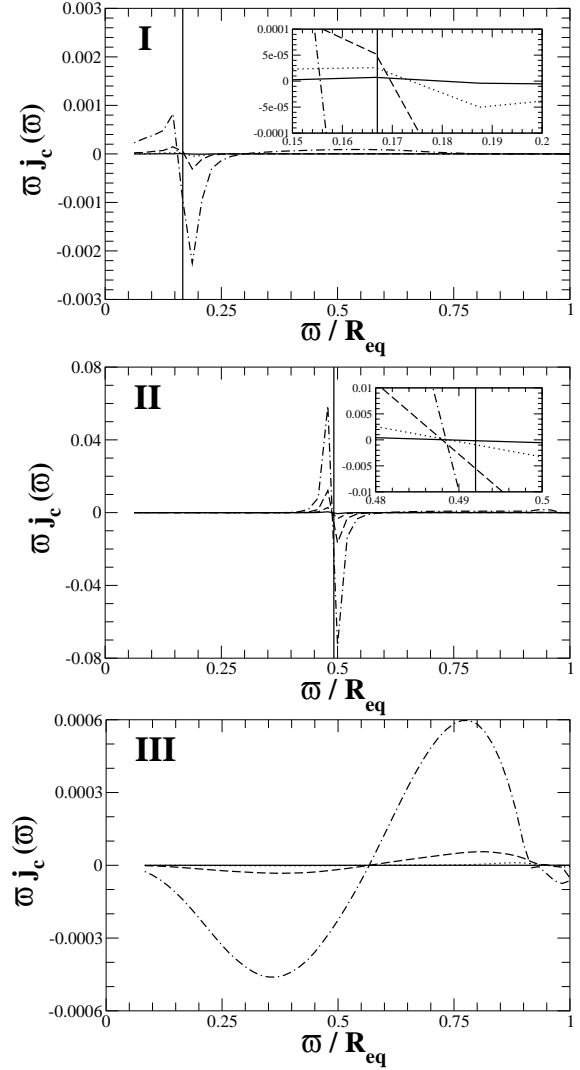}
\caption{
Snapshots of the canonical angular momentum distribution $\varpi
j_{\rm c} (\varpi)$ in the equatorial plane for three differentially
rotating stars (see Table~\ref{tab:initial}).  Solid, dotted, dashed,
and dash-dotted line represents the time
$t/P_{\rm c} =$(3.47, 6.93, 10.40, 13.86) for model~I,
$t/P_{\rm c} =$(45.68, 56.43, 67.18, 77.97) for model~II, and 
$t/P_{\rm c} =$(1.10, 2.19, 3.29, 4.39) for model~III, respectively.
Note that vertical line in panels~I and II denotes the corotation
radius of the star (model~III does not have a corotation radius).  We
also enlarged the figure around the corotation radius for panels~I and
II, which is presented in the right upper part of each panel.
Although our method of determining the corotation radius is not
precise, we clearly find that the distribution significantly changes
around the corotation radius.
}
\label{fig:jc}
\end{figure}

We show the snapshots of canonical angular momentum density in 
Fig.~\ref{fig:jc}.  Since we determine the corotation radius using the 
extracted eigenfrequency and the angular velocity profile at 
equilibrium, the radius does not change throughout the evolution.  For
low $T/|W|$ dynamical instability, the distribution of the canonical
angular momentum drastically changes its sign around the corotation
radius, and the maximum amount of canonical angular momentum density
increases at the early stage of evolution.  This means that the
angular momentum flows inside the corotation radius in the evolution.
On the other hand, for high $T/|W|$ dynamical instability (the bottom
panel of Fig.~\ref{fig:jc}), which may be regarded as a classical 
bar instability modified by differential rotation, the distribution of the 
canonical angular momentum is smooth and with no particular feature. 

Note that the amplitude of $\varpi j_c$ is orders of magnitude
smaller than those in the corotating cases in top and middle panels of
Fig.~\ref{fig:jc}.  Contrary to the linear perturbation analysis in
Section~\ref{subsec:AppCylinder}, the amplitude here is not scale free
and the relative amplitude has a physical meaning.  Thus the smallness
of it for model~III suggest that it should be exactly zero everywhere
in the limit of linearized oscillation.  The deviation from zero may
come from the imperfect assumption of linearized oscillation, that is
made here to extract oscillation frequency and Lagrangian displacement
vector. 

From these different behaviours of the distribution of the canonical
angular momentum, we find that the mechanisms working in the low
$T/|W|$ instabilities and the classical bar instability may be quite
different, i.e., in the former the corotation resonance of modes are
essential, while the instability is global in the latter case.

\section{Summary and Discussion}
\label{sec:Discussion}

We have studied the nature of three different types of dynamical
instability in differentially rotating stars both in linear eigenmode
analysis and in hydrodynamic simulation using canonical angular
momentum distribution.

We have found that the low $T/|W|$ dynamical instability occurs around
the corotation radius of the star by investigating the distribution of
the canonical angular momentum.  We have also found by investigating
the canonical angular momentum that the instability grows through the
inflow of the angular momentum inside the corotation radius.  The
feature also holds for the dynamical bar instability in low $T/|W|$,
which is in clear contrast to that of classical dynamical bar
instability in high $T/|W|$.  Therefore the existence of corotation
point inside the star plays a significant role of exciting one-armed
spiral mode and bar mode dynamically in low $T/|W|$.  However, we made
our statement from the behaviour of the canonical angular momentum,
the statement holds only in a sense of necessary condition.  In order
to understand the physical mechanism of the low $T/|W|$ dynamical
instability, we need another tool and it will be the next step of this
study.

The feature of gravitational waves generated from these instabilities
are also compared.  Quasi-periodic gravitational waves emitted by
stars with $m=1$ instabilities have smaller amplitudes than those
emitted by stars unstable to the $m=2$ bar mode.  For $m=1$ modes, the
gravitational radiation is emitted not directly by the primary mode
itself, but by the $m=2$ secondary harmonic which is simultaneously
excited.  Possibly this $m=2$ oscillation is generated through a
quadratic nonlinear selfcoupling of $m=1$ eigenmode.  Remarkably the
precedent studies \citep{CNLB01,SBS03} found that the pattern speed of
$m=2$ mode is almost the same as that of $m=1$ mode, which suggest the
former is the harmonic of the latter.  Unlike the case for
bar-unstable stars, the gravitational wave signal does not persist for
many periods, but instead is damped fairly rapidly.  We have not
understood this remarkable difference between $m=1$ and $m=2$ unstable
cases.  One of the possibility may be that the unstable $m=1$
eigenmode tends to couple to higher and higher $m$ modes (which are
not necessarily unstable and could be in the continuous spectrum) and
pump its energy to them in a cascade way.  However, we have not found
the feature that prevents $m=2$ mode from this cascade dissipation.

Another possibility is that the spiral pattern formed in $m=1$
instability redistributes the angular momentum of the original
unstable flow, so that the flow is quickly stabilized.  Inside the
corotation radius, the background flow is faster than the pattern,
while it is slower outside.  A similar mechanism to Landau damping in
plasma wave which transfer the momentum of wave to background flow may
work at the spiral pattern.  The pattern may decelerate the background
flow inside the corotation and accelerate it outside the corotation,
which may change the unstable flow profile to stable one.  As we do
not see a spiral pattern forming in the low $T/|W|$ bar instability,
it may eludes this damping process.

\section*{Acknowledgements}
We are grateful to John Friedman for fruitful discussions and useful 
suggestions.  We thank Anna Watts for useful comments on our work and
for her unpublished material on canonical energy of rotating shell
model, and Charles Gammie for his suggestion on relevant literatures.
We also thank Takashi Nakamura, Misao Sasaki, Takahiro Tanaka for
stimulating discussions and comments.  
We furthermore thanks an anonymous referee for his/her careful reading 
of our manuscript.
This work was supported in part
by MEXT Grant-in-Aid for young scientists (No. 200200927), by the PPARC 
grant (PPA/G/S/2002/00531) at the University of Southampton, and by NSF
grant (PHY-0071044).  Numerical computations were performed on the
VPP-5000 machine in the Astronomical Data Analysis Centre, National
Astronomical Observatory of Japan, on the Pentium-4 cluster machine in
the Theoretical Astrophysics Group at Kyoto University.


\appendix
\section{Canonical angular momentum of classical bar instability}
\label{appendix-classical}
In this appendix we evaluate the angular momentum density of bar
instability in Bardeen disk and in Maclaurin spheroid.  This quantity
is defined in the cylindrical coordinate as
\begin{eqnarray}
j_c(\varpi) &=& m\Re[s]\rho|\xi|^2 - 2m\Omega\rho
 ~\Im[\xi^\varpi\cdot\xi^{\varphi*}]
\label{jc1}\\
&=& \frac{m\rho}{|s|^2}\left(\Re[s]|\delta v|^2+2\Omega~\Im\left[
\delta v^\varpi\cdot \delta v^{\varphi*}\right]\right),
\label{jc2}
\end{eqnarray}
where $s=\sigma-m\Omega$ is the frequency of the mode observed in
corotating frame of the fluid.  The components of Lagrangian
displacement $\xi$ and Eulerian velocity $\delta v$ are written in the
local orthonormal frame.

\subsection{Bardeen disk}
\citet{Bardeen75} analytically constructed a self-gravitating `warm'
disk with finite thickness by finding corrections to an
infinitesimally thin disk from which he started his approximation.
Perturbation of a Bardeen disk is studied in \citet{SB83}.  Here, we
follow their definitions and notations.  Velocity perturbation is
written with two potential functions $\alpha$ and $\beta$ as
\begin{equation}
\delta v = -i\nabla\alpha - \nabla\times(\beta e_z),
\end{equation}
where $e_z$ is the unit vector in $z$-direction.  Introducing $\eta = 
\sqrt{1-({\varpi}/{R})^2}$ in terms of cylindrical radial coordinate
$\varpi$, the general perturbation can be expanded by Legendre
polynomials as
\begin{equation}
\alpha = \sum_{l=0}^{\infty} \sum_{m=-\infty}^{\infty}
\alpha_l^m P_{2l+m}^m(\eta) e^{im\varphi}.
\end{equation}
Master equations (equations~(3.1) and (3.4) in \citet{SB83}) for
perturbation can be decomposed into those for each $l$ and $m$.  The
bar mode is $l=0$, $m=0$ case.  From equations~(3.13) to (3.15) of
\citet{SB83}, we get a simple eigenvalue equation 
\begin{equation}
s^2 - 2\Omega s 
+ \frac{3}{2}-\frac{16\mu}{\pi}= 0,
\end{equation}
where $\Omega$ is the angular velocity of disk 
\begin{equation}
\Omega = \sqrt{1-\frac{8\mu}{\pi}},
\end{equation}
while $s$ is a mode frequency observed in a corotating frame with
disk.  Both of them are normalized by $\Omega_{\rm cld}$, an angular
frequency of cold disk.  $\mu$ is called `aspect ratio' parameter
(equation~(2.18) in \citet{SB83}).  When this parameter is smaller
than ${\pi}/{16}$, one of the modes above becomes dynamically
unstable.  The corresponding radial eigenfunctions are written as
\begin{eqnarray}
\alpha_2^2 &=& K
\left(1+\frac{8\Gamma}{s(s+2\Omega)}\right)P_2^2({\eta}), 
\\
\beta_2^2 &=& K
\left(1-\frac{8\Gamma}{s(s+2\Omega)}\right)P_2^2({\eta}), 
\end{eqnarray}
where 
\begin{equation}
\Gamma = \frac{3}{16}-\frac{2\mu}{\pi}. 
\end{equation}
Using $\alpha_2^2$ and $\beta_2^2$,Eulerian perturbation of velocity
for bar mode is written as
\begin{eqnarray}
\delta v^\varpi &=& -i\left(\frac{d}{d\varpi}\alpha_2^2
+\frac{2}{\varpi}\beta_2^2\right) = -6i K\varpi,
\\
\delta v^\varphi &=& \frac{d}{d\varpi}\beta_2^2
+ \frac{2}{r}\alpha_2^2 = 6K \varpi,
\end{eqnarray}
where we used the definition of $\eta$.  With bearing in mind that the
real part of $s$ is $\Omega$, equation~(\ref{jc2}) gives that $j_c$ is
0.

\subsection{Maclaurin spheroid}
For the bar mode of Maclaurin spheroid, we have an analytical
expression of Lagrangian displacement,
\begin{equation}
\xi^x = k\varpi e^{\pm i\varphi},~
\xi^y = k\varpi e^{\pm i(\varphi+\frac{\pi}{2})},~
\xi^z = 0,
\end{equation}
in Cartesian components \citep{Chandra69}.  Here $k$ is a constant.
It is straightforward to see that the corresponding components in
cylindrical coordinate are
\begin{equation}
\xi^\varpi = k\varpi e^{\pm 2i\varphi},~
\xi^\varphi = \pm k e^{\pm 2i\varphi},~
\xi^z = 0.
\end{equation}
$\xi^\varphi$ is defined with respect to coordinate base vector
$\partial / \partial\varphi$.  From equation~(\ref{jc1}) with the fact
that real part of frequency of unstable bar mode is $\Omega$, we
easily see that $j_{\rm c}$ vanishes everywhere.
\label{lastpage}

\begin{thebibliography}{99}
%
\bibitem[\protect\citeauthoryear{Acheson}{1976}]{Acheson76}
Acheson D. J., 1976, 
J. Fluid Mech., 77, 433
%
\bibitem[\protect\citeauthoryear{Balbinski}{1985}]{Balbinski85}
Balbinski E., 1985, 
MNRAS, 216, 897
%
\bibitem[\protect\citeauthoryear{Bardeen}{1975}]{Bardeen75}
Bardeen J. M., 1975, 
in Hayli A., eds., Proc. IAU Symp. 69,
Dynamics of Stellar Systems.
Reidel, Dordrecht, p. 297
%
\bibitem[\protect\citeauthoryear{Blaes}{1985a}]{Blaes85a}
Blaes O. M., 1985a, 
MNRAS, 212, 37
%
\bibitem[\protect\citeauthoryear{Blaes}{1985b}]{Blaes85b}
Blaes O. M., 1985b, 
MNRAS, 216, 553
%
\bibitem[\protect\citeauthoryear{Blaes \& Glatzel}{1986}]{Blaes86}
Blaes O. M., Glatzel W., 1986, 
MNRAS, 220, 253
%
\bibitem[\protect\citeauthoryear
{Cook, Shapiro \& Stephans}{Cook et al.}{2003}]{CSS03}
Cook J. N., Shapiro S. L., Stephens B. C., 2003, 
ApJ, 599, 1272
%
\bibitem[\protect\citeauthoryear{Centrella et al.}{2001}]{CNLB01}
Centrella J. M., New K. C. B., Lowe L. L., Brown, J. D., 2001,
ApJ, 550, L193
%
\bibitem[\protect\citeauthoryear{Chandrasekhar}{1969}]{Chandra69}
Chandrasekhar S., 1969,
Ellipsoidal Figures of Equilibrium.
Yale Univ. Press, New York, Chap. 5, Sec. 33
%
\bibitem[\protect\citeauthoryear
{Christodoulou \& Narayan}{1992}]{CN92}
Christodoulou D. M., Narayan R., 1992, 
ApJ, 388, 451
%
\bibitem[\protect\citeauthoryear{Drury}{1985}]{Drury85}
Drury, L. O'C., 1985, 
MNRAS, 217, 821
%
\bibitem[\protect\citeauthoryear
{Duez, Shapiro \& Yo}{Duez et al.}{2004}]{DSY04}
Duez M. D., Shapiro S. L., Yo H.-J., 2004,
PRD 69, 104016
%
\bibitem[\protect\citeauthoryear{Durisen et al.}{1986}]{DGTB86}
Durisen R. H., Gingold R. A., Tohline J. E., Boss A. P., 1986, 
ApJ, 305, 281
%
\bibitem[\protect\citeauthoryear{Friedman \& Schutz}{1978a}]{FS78a}
Friedman J. L., Schutz B. F., 1978a,
ApJ, 221, 937
%
\bibitem[\protect\citeauthoryear{Friedman \& Schutz}{1978b}]{FS78b}
Friedman J. L., Schutz B. F., 1978b,
ApJ, 222, 281
%
\bibitem[\protect\citeauthoryear{Glatzel}{1987a}]{Glatzel87a}
Glatzel W., 1987a, 
MNRAS, 225, 227
%
\bibitem[\protect\citeauthoryear{Glatzel}{1987b}]{Glatzel87b}
Glatzel W., 1987b, 
MNRAS, 228, 77
%
\bibitem[\protect\citeauthoryear
{Goldreich, Goodman \& Narayan}{Goldreich et al.}{1986}]{GGN86}
Goldreich P., Goodman J., Narayan R., 1986, 
MNRAS, 221, 339
%
\bibitem[\protect\citeauthoryear{Goodman \& Narayan}{1988}]{GN88}
Goodman J., Narayan R., 1988, 
MNRAS, 231, 97
%
\bibitem[\protect\citeauthoryear{Hachisu}{1986}]{Hachisu86}
Hachisu I., 1986,
ApJS, 61, 479
%
\bibitem[\protect\citeauthoryear{Houser \& Centrella}{1996}]{HC}
Houser J. L., Centrella J. M., 1996, 
PRD, 54, 7278
%
\bibitem[\protect\citeauthoryear
{Houser, Centrella \& Smith}{Houser et al.}{1994}]{HCS}
Houser J. L., Centrella J. M., Smith S. C., 1994, 
PRL 72, 1314
%
\bibitem[\protect\citeauthoryear{Ishibashi \& Ando}{1985}]{IA85}
Ishibashi S., Ando H., 1985, 
PASJ, 37, 97
%
\bibitem[\protect\citeauthoryear{Ishibashi \& Ando}{1986}]{IA86}
Ishibashi S., Ando H., 1986, 
PASJ, 38, 295
%
\bibitem[\protect\citeauthoryear{Kojima}{1986}]{Kojima86}
Kojima Y., 1986, 
Prog. Theor. Phys., 75, 251
%
\bibitem[\protect\citeauthoryear{Kojima}{1989}]{Kojima89}
Kojima Y., 1989, MNRAS, 236, 589
%
\bibitem[\protect\citeauthoryear{Luyten}{1988}]{Luyten88}
Luyten P., 1988, 
Ap\&SS, 141, 27
%
\bibitem[\protect\citeauthoryear{Luyten}{1989}]{Luyten89}
Luyten P., 1989, 
Ap\&SS, 153, 13
%
\bibitem[\protect\citeauthoryear{Luyten}{1990}]{Luyten90}
Luyten P., 1990, 
MNRAS, 242, 447
%
\bibitem[\protect\citeauthoryear
{Narayan, Goldreich \& Goodman}{Narayan et al.}{1987}]{NGG87}
Narayan R., Goldreich P., Goodman J., 1987, 
MNRAS, 228, 1
%
\bibitem[\protect\citeauthoryear
{Misner, Thorne \& Wheeler}{Misner et al.}{1973}]{MTW}
Misner C. W., Thorne K. S., Wheeler J. A., 1973, 
Gravitation. Freeman, New York, Chap. 36.10
%
\bibitem[\protect\citeauthoryear
{New, Centrella \& Tohline}{New et al.}{2000}]{NCT}
New K. C. B., Centrella J. M., Tohline J. E., 2000,
PRD, 62, 064019
%
\bibitem[\protect\citeauthoryear{Ostriker}{1965}]{Ostriker65}
Ostriker J., 1965, 
ApJS, 11, 167
%
\bibitem[\protect\citeauthoryear{Ott et al.}{2005}]{Ott05}
Ott C. D., Ou S., Tohline J. E., Burrows A., 2005, 
ApJL, 625, 119
%
\bibitem[\protect\citeauthoryear{Papaloizou \& Pringle}{1984}]{PP84}
Papaloizou J. C. B., Pringle J. E., 1984,
MNRAS, 208, 721
%
\bibitem[\protect\citeauthoryear{Papaloizou \& Pringle}{1985}]{PP85}
Papaloizou J. C. B., Pringle J. E., 1985,
MNRAS, 213, 799
%
\bibitem[\protect\citeauthoryear{Papaloizou \& Pringle}{1987}]{PP87}
Papaloizou J. C. B., Pringle J. E., 1987,
MNRAS, 225, 267
%
\bibitem[\protect\citeauthoryear
{Pickett, Durisen \& Davis}{Pickett et al.}{1996}]{PDD}
Pickett B. K., Durisen R. H., Davis G. A., 1996,
ApJ, 458, 714
%
\bibitem[\protect\citeauthoryear{Robe}{1979}]{Robe79}
Robe H., 1979, 
A\&A, 75, 14
%
\bibitem[\protect\citeauthoryear{Ruoff \& Kokkotas}{2001}]{RK01}
Ruoff J., Kokkotas K. D., 2001, 
MNRAS, 328, 678
%
\bibitem[\protect\citeauthoryear{Saijo}{2005}]{Saijo05}
Saijo M., 2005,
PRD, 71, 104038
%
\bibitem[\protect\citeauthoryear{Saijo et al.}{2001}]{SSBS}
Saijo M., Shibata M., Baumgarte T. W., Shapiro S. L., 2001,
ApJ, 548, 919
%
\bibitem[\protect\citeauthoryear
{Saijo, Baumgarte \& Shapiro}{Saijo et al.}{2003}]{SBS03}
Saijo M., Baumgarte T. W., Shapiro S. L., 2003,
ApJ, 595, 352
%
\bibitem[\protect\citeauthoryear{Schutz \& Bowen}{1983}]{SB83}
Schutz B. F., Bowen A. M., 1983, 
MNRAS, 202, 867
%
\bibitem[\protect\citeauthoryear{Shapiro \& Teukolsky}{1983}]{ST83}
Shapiro S. L., Teukolsky S. A., 1983, 
Black Holes, White Dwarfs, and Neutron Stars. 
John Wiley and Sons, New York, Chap. 7.5
%
\bibitem[\protect\citeauthoryear
{Shibata, Baumgarte \& Shapiro}{Shibata et al.}{2000}]{SBS00}
Shibata M., Baumgarte T. W., Shapiro S. L., 2000,
ApJ, 542, 453
%
\bibitem[\protect\citeauthoryear
{Shibata, Karino \& Eriguchi}{Shibata et al.}{2002}]{SKE02}
Shibata M., Karino S., Eriguchi Y., 2002,
MNRAS, 334, L27
%
\bibitem[\protect\citeauthoryear
{Shibata, Karino \& Eriguchi}{Shibata et al.}{2003}]{SKE03}
Shibata M., Karino S., Eriguchi Y., 2003,
MNRAS, 343, 619
%
\bibitem[\protect\citeauthoryear
{Shibata \& Sekiguchi}{2005}]{SS05}
Shibata M., Sekiguchi Y.-I., 2005, 
PRD, 71, 024014
%
\bibitem[\protect\citeauthoryear{Shu}{1992}]{Shu92}
Shu F. H., 1992, 
The Physics of Astrophysics: Volume II Gas Dynamics.
University Science Book, Mill Valley, California, Chapters 11, 12
%
\bibitem[\protect\citeauthoryear
{Smith, Houser \& Centrella}{Smith et al.}{1995}]{SHC}
Smith S. C., Houser J. L., Centrella J. M., 1995, 
ApJ, 458, 236
%
\bibitem[\protect\citeauthoryear{Stix}{1992}]{Stix92}
Stix T. H., 1992, 
Waves in Plasmas. 
American Institute of Physics, New York, Chap. 8
%
\bibitem[\protect\citeauthoryear{Tassoul}{1978}]{Tassoul78} 
Tassoul J.-L., 1978, Theory of Rotating Stars.
Princeton Univ. Press., Princeton, Chap. 10
%
\bibitem[\protect\citeauthoryear
{Tohline, Durisen \& McCollough}{Tohline et al.}{1985}]{TDM}
Tohline J. E., Durisen R. H., McCollough M., 1985,
ApJ, 298, 220
%
\bibitem[\protect\citeauthoryear{Tohline \& Hachisu}{1990}]{TH90}
Tohline J. E., Hachisu I., 1990, 
ApJ, 361, 394 
%
\bibitem[\protect\citeauthoryear{Toman et al.}{1998}]{TIPD}
Toman J., Imamura J. N., Pickett B. J., Durisen R. H., 1998,
ApJ, 497, 370
%
\bibitem[\protect\citeauthoryear{Wald}{1984}]{Wald84}
Wald R. M., 1984,
General Relativity. Univ. Chicago Press, Chicago,
Appendix E.
%
\bibitem[\protect\citeauthoryear{Watts et al.}{2003}]{WABS03}
Watts A. L., Andersson N., Beyer H., Schutz B. F., 2003, 
MNRAS, 342, 1156
%
\bibitem[\protect\citeauthoryear
{Watts, Andersson \& Jones}{Watts et al.}{2005}]{WAJ05}
Watts A. L., Andersson N., Jones D. I., 2005, 
ApJL, 618, 37
%
\bibitem[\protect\citeauthoryear{Williams \& Tohline}{1988}]{WT}
Williams H. A., Tohline J. E., 1988, 
ApJ, 334, 449
%
\bibitem[\protect\citeauthoryear{Zink et al.}{2005}]{Zink05}
Zink B., Stergioulas N., Hawke I., Ott C. D., Schnetter E., 
M\"{u}ller E., 2005, preprint (gr-qc/0501080)
%
\end{thebibliography}
\end{document}